\documentclass[12pt]{amsart}
\usepackage{amsfonts,amssymb,amsmath,amscd}
\usepackage{epsf}

\newcommand{\ClMod}{{\mathsf {CLMOD}}}
\newcommand{\K}{{\mathsf K}}

\newcommand{\bj}{{\bar{j}}}

\newcommand{\sm}{\sigma}

\renewcommand{\b}{{\beta}}
\renewcommand{\a}{{\alpha}}
\newcommand{\la}{{\lambda}}
\renewcommand{\L}{{\mathcal L}}
\newcommand{\N}{{\mathcal N}}

\renewcommand{\d}{{\rm d}}

\newcommand{\ep}{{\epsilon}}
\newcommand{\bep}{{\bar\ep}}
\newcommand{\ga}{\gamma}
\newcommand{\om}{\omega}

\renewcommand{\l}{\left}
\renewcommand{\r}{\right}

\newcommand{\ta}{{\tilde{a}}}

\newcommand{\taa}{{\tilde{\a}}}

\def\tbb{{\tilde{\b}}}

\newcommand{\LG}{Landau-Ginzburg }
\newcommand{\C}{{\mathsf C}}

\newcommand{\vt}{\vspace{2mm}}

\newcommand{\Tr}{{\rm Tr}}

\renewcommand{\th}{{\theta}}

\newcommand{\CC}{{\mathbb C}}
\newcommand{\PP}{{\mathbb P}}
\newcommand{\RR}{{\mathbb R}}
\newcommand{\ZZ}{{\mathbb Z}}
\newcommand{\NN}{{\mathbb N}}
\renewcommand{\SS}{{\mathbb S}}

\newcommand{\End}{{\rm End}}

\newcommand{\ra}{\rightarrow}
\newcommand{\bpartial}{{\bar\partial}}
\newcommand{\no}{\nonumber}
\newcommand{\op}{\oplus}
\newcommand{\ot}{\otimes}
\renewcommand{\O}{{\mathcal O}}
\newcommand{\B}{{\mathcal B}}

\title[D-Branes in LG Models and Algebraic Geometry]{D-Branes in 
Landau-Ginzburg Models and Algebraic Geometry}
\author[A. Kapustin]{Anton Kapustin \hspace{3mm} }
\address{California Institute of Technology\\
Department of Physics\\
Pasadena, CA 91125}
\email{kapustin@theory.caltech.edu}
\author[Y. Li]{ \hspace{3mm} Yi Li}
\address{California Institute of Technology\\
Department of Physics\\
Pasadena, CA 91125}
\email{yili@theory.caltech.edu}

\date{August 2002}

\begin{document}

\begin{abstract}
We study topological D-branes of type B in $N=2$ Landau-Ginzburg 
models, focusing on the case where all vacua have a mass gap. 
In general, tree-level topological 
string theory in the presence of topological D-branes is described 
mathematically in terms of a triangulated category. For example, it has
been argued that B-branes for an $N=2$ sigma-model with a Calabi-Yau target 
space are described by the derived category of coherent sheaves on this
space. M.~Kontsevich previously proposed a candidate category for 
B-branes in $N=2$ Landau-Ginzburg models, and our computations 
confirm this proposal. We also give a heuristic physical derivation of the proposal.
Assuming its validity, we can completely
describe the category of B-branes in an arbitrary massive 
Landau-Ginzburg model in terms of modules over a Clifford algebra.
Assuming in addition Homological Mirror Symmetry, our results enable one to compute the Fukaya category
for a large class of Fano varieties. 
We also provide a (somewhat trivial) counter-example to the hypothesis that given a closed
string background there is a unique set of D-branes consistent with it.

\end{abstract}

\maketitle

\vspace{-5in}

\parbox{\linewidth}
{\small\hfill \shortstack{CALT-68-2412}} 

\vspace{5in}

\section{Introduction}

Topological open strings and topological D-branes have recently been enjoying the attention of both physicists and mathematicians.
The most obvious physical
motivation for studying topological string theory  is that it is a toy-model for ``physical'' string theory. 
Thus a better understanding of topological D-branes could shed light
on the general definition of a boundary condition for a two-dimensional conformal field theory (2d CFT), something
which is not known at present. Further, if a 2d topological field theory (2d  TFT) is obtained by twisting a 2d supersymmetric 
field theory, then it is possible to regard topological D-branes as a special class of ``physical'' D-branes (BPS D-branes).
In fact, much of recent progress in string theory has resulted from studying BPS D-branes.

{}From the
mathematical viewpoint, topological string theory is an alternative
way of describing certain important geometric categories, such as the
category of coherent sheaves on a Calabi-Yau manifold, and can serve as a
powerful source of intuition. An outstanding example of such intuition is the
Homological Mirror Symmetry conjecture~\cite{K1}.

Most works on topological string theory considered the case of topologically
twisted $N=2$ sigma-models~\cite{W:sm} with a Calabi-Yau target space. This is the
case when the world-sheet theory is conformal, and topological correlators 
can also be interpreted in terms of a physical string theory~\cite{W:mir}. However, 
one can also consider more general topologically twisted $N=2$
field theories and the corresponding D-branes. One class of such theories
is given by sigma-models whose target is a Fano variety (say, a complex
projective space, or a complex Grassmannian). Such QFTs, although
conformally-invariant on the classical level, have non-trivial renormalization-group
flow once quantum effects are taken into account. Another set of examples 
is provided by $N=2$ Landau-Ginzburg models (LG models)~\cite{V}. In fact, in many cases
these two classes of $N=2$ theories are related by mirror symmetry~\cite{HV}.
For example, the sigma-model with target $\CC\PP^n$ is mirror to a 
Landau-Ginzburg model with $n$ fields $x_1,\ldots, x_n,$  taking values
in $\CC^*$, and a superpotential
\begin{equation}\label{WCPn}
W=x_1+\ldots +x_n +\frac{1}{x_1\ldots x_n}.
\end{equation}
Thus if one wants to extend the Homological Mirror Symmetry conjecture
to the non-Calabi-Yau case, one needs to understand D-branes in 
topologically twisted LG models. Note that all critical points
of this superpotential are isolated and non-degenerate; this means that
all the vacua have a mass gap, and the infrared limit of this LG model
is trivial. In what follows we will call such LG models {\it massive}. 
Despite the triviality of the infrared limit, the Homological Mirror 
Symmetry conjecture remains meaningful and non-trivial in this case.

Very recently it has been proposed that massive $N=2$ $d=2$ QFTs can be used
to describe certain non-standard superstring backgrounds with Ramond-Ramond flux~\cite{ML}.
Thus a study of D-branes in massive QFTs could be useful for understanding open strings
in such Ramond-Ramond backgrounds.

In order to formulate our problem more concretely, let us first
summarize the situation in the Calabi-Yau case, where the $N=2$ field theory
is conformal. $N=2$ superconformal field theories
have two topologically twisted versions: A-model and B-model~\cite{W:sm,W:mir,EY}. The
corresponding D-branes are called A-branes and B-branes. 
Mirror symmetry exchanges A-branes and B-branes. Tree-level topological correlators
give the set of either A-branes or B-branes the structure of an 
$A_\infty$-category; gauge-invariant information is encoded by the 
corresponding derived categories. It has been argued that the
derived category of B-branes is equivalent to the derived category
of coherent sheaves~\cite{K1,Douglas,AL}. A detailed check of this proposal has been
performed in Ref.~\cite{KatzSharpe}.

For A-branes on Calabi-Yau 
manifolds, it has been proposed that the relevant $A_\infty$-category is the so-called 
Fukaya category, 
whose objects are (roughly) Lagrangian submanifolds carrying vector 
bundles with flat connections~\cite{K1}. Recently it has 
been shown that the derived Fukaya category is too small and does not accommodate 
certain physically acceptable A-branes~\cite{KO}. In particular, if we want the
Homological Mirror Symmetry conjecture to be true for tori, then the
Fukaya category must be enlarged with non-Lagrangian (more specifically,
coisotropic) branes.

In the case of Fano varieties, the sigma-model is not conformal. What is more important,
the axial $U(1)$ R-current is anomalous, and therefore one cannot define the B-twist~\cite{W:mir}.
One {\it can} consider D-branes which preserve B-type supersymmetry, but the relation with the derived
category of coherent sheaves is less straightforward~\cite{HIV,Hbook}. 
Mirror symmetry relates B-branes on Fano varieties with A-branes in LG models. The latter have been
studied from a variety of viewpoints in Refs.~\cite{GJS,GJ,HIV,H}. 
In the case when the Fano variety is $\CC\PP^n$, the prediction of mirror symmetry
has been tested in Ref.~\cite{HIV,Seidel}. In particular, the mirrors of ``exceptional'' bundles
on $\CC\PP^n$ have been identified, and in the case of $\CC\PP^1$ and $\CC\PP^2$ it has
been checked that morphisms between these bundles
in the derived category of coherent sheaves on $\CC\PP^n$
agree with the Floer homology between their mirror A-branes.

One can also consider the category of A-branes on a Fano manifold. Since the vector $U(1)$
R-current is not anomalous, the A-twist is well-defined, and A-branes can be regarded as topological
boundary conditions for the A-model. 
Presumably, the category of A-branes contains the derived Fukaya category as a subcategory,
but other than that little is known about it, even in the case of
$\CC\PP^n$. If we assume mirror symmetry, we can learn about the category of 
A-branes on $\CC\PP^n$ by studying B-branes in the mirror LG model. The B-twist is well-defined
for any LG model whose target has a trivial canonical bundle, thus B-branes in such a LG model can
be regarded as topological boundary conditions for the B-model. An obvious question is how the
introduction of the superpotential deforms the relation between the category of B-branes and the
derived category of coherent sheaves.

Important steps towards understanding B-branes in LG models have been taken in Refs.~\cite{GJS,HIV,Hbook,H,HICM}
(see also Refs.~\cite{GJ2,HMcG,Hetal} for a related work). 
In these papers general properties of B-branes have been studied, and several concrete examples
have been discussed. A somewhat surprising lesson from these works is that the category of
B-branes remains non-trivial even in a massive LG model, where the bulk 2d TFT is trivial. For example,
if we take the superpotential $W$ to be a non-degenerate quadratic function on $\CC^n$,
the graded algebra of endomorphisms of a 
D0-brane sitting at the critical point of $W$ is isomorphic to a Clifford
algebra with $n$ generators~\cite{H}. This raises the question if one can determine
the category of B-branes in any massive LG model.

A proposal which accomplishes this has been put forward by M.~Kontsevich.
Roughly speaking, the proposal is that the superpotential $W$ 
deforms the derived category of coherent sheaves by replacing complexes of 
locally free sheaves with ``twisted'' complexes. Here ``twisted'' means
that compositions of successive morphisms in a complex are equal to $W$, 
instead of zero.
One also needs to switch from $\ZZ$-graded complexes to $\ZZ_2$-graded ones.
Kontsevich's proposal is supposed to describe B-branes in any LG model
such that the critical set of $W$ is compact; in particular, it does not
require the critical points of $W$ to be non-degenerate. 

The main goal of this paper is to provide evidence for Kontsevich's proposal. Our evidence is
of two kinds. First, we argue on physical grounds that twisted complexes arise as a consequence of
BRST-invariance. More precisely, while in the presence of the
superpotential a holomorphic vector bundle or a complex of vector bundles does not correspond to
a B-type boundary condition, we show that any twisted
complex of holomorphic vector bundles is a valid B-brane. Second, we test the proposal in some specific
cases where morphisms between branes (i.e. spectra of topological open strings) can be easily computed.
We focus on the massive case,
where the proposal simplifies considerably. Namely, the category of
$\ZZ_2$-graded twisted complexes can be related to the sum of several copies of
the category of finite-dimensional $\ZZ_2$-graded
modules over a Clifford algebra. We will denote the latter category $\C_{tot}$ in what follows.
This reformulation is helpful,
because the functor from the category of B-branes in a massive LG model
to $\C_{tot}$ is very simple to describe. In this paper 
we perform some checks that this functor is an embedding of graded 
categories. In view of the above-mentioned ``duality'' between $\ZZ_2$-graded
twisted complexes and $\C_{tot}$, this provides a
test of Kontsevich's proposal. Assuming the validity of the proposal, we infer that the category of B-branes
in an arbitrary massive LG model is a full sub-category of $\C_{tot}$. 
The latter has a very simple and explicit description.

Since for many Fano varieties the mirror LG model is known, our results allow one to
effectively compute the category of A-branes for such varieties. From the mathematical
viewpoint, it is an interesting challenge to reproduce such results using methods
of symplectic geometry.

Axiomatic definitions of topological D-branes for 2d Topological Field Theories (2d TFTs)
have been recently proposed by G.~Moore and G.~Segal~\cite{MS}
and C.~I.~Lazaroiu~\cite{Laz}. One of the main unresolved problems in the 
axiomatic approach is whether these axioms determine unambiguously the
category of topological D-branes associated to a given TFT. We show
that the category of B-branes for a massive LG model with a quadratic 
superpotential provides a counter-example to uniqueness. In fact, this
example shows that uniqueness, if understood naively, fails also for 
ordinary (i.e. non-topological) D-branes in any closed superstring
background. However, this particular failure is rather mild, i.e.
it does not seem to have serious physical consequences.

Now let us describe the content of the paper in more detail. 
In Section~\ref{sec:A} we recall some basic facts about mirror symmetry
between Fano varieties and LG models. In Section~\ref{sec:B} we 
review general properties of B-branes in LG models. We argue
that if the superpotential has only isolated critical points, then
it is sufficient to study B-branes in the infinitesimal neighborhood of each
critical point. For example, if all critical points of $W$ are non-degenerate,
one does not lose anything if one replaces the superpotential by
its quadratic approximation near each critical point. The material in this section is
not new and has been previously discussed in Refs.~\cite{HIV,Hbook,H,HICM}.
In Sections~\ref{sec:C} and \ref{sec:Cp} we study B-branes in the LG model 
with the superpotential $W=xy.$ This is the simplest LG model where B-branes 
of dimension larger than $0$ are present. In Section~\ref{sec:D} we 
discuss B-branes in more general LG models with the superpotential
$W=z_1^2+z_2^2+\ldots+z_n^2.$ In Section~\ref{sec:E} we explain Kontsevich's
proposal and show that our results are consistent with it. We also explain why BRST
invariance of boundary conditions requires twisted complexes of vector bundles
instead of ordinary complexes.
This provides a physical explanation of Kontsevich's proposal. We also
relate B-branes in massive LG models to $\ZZ_2$-graded Clifford modules.
In Section~\ref{sec:F} we use Homological Mirror Symmetry to compute the category of
A-branes for $\CC\PP^2$ and $\CC\PP^1\times\CC\PP^1$.
Section~\ref{sec:G} contains concluding remarks.

\section{Mirror Symmetry for Fano varieties and LG models}
\label{sec:A}

A Fano variety is a compact complex manifold whose anti-canonical line bundle
is ample. This is equivalent to saying that the first Chern class of the
canonical line bundle is
negative-definite. An $N=2$ sigma-model whose target space is a Fano variety
describes an $N=2$ $d=2$ field theory which is free in the ultraviolet. In the
infrared, it can either flow to a massive vacuum, or to a non-trivial $N=2$ SCFT.
Note that classically $N=2$ sigma-models have both vector and axial $U(1)$ 
R-symmetries, but for Fano varieties quantum anomalies break the axial 
R-symmetry down to a discrete subgroup. Generically, this subgroup is $\ZZ_2$, but in
special cases it can be larger. In the Calabi-Yau case
the full axial R-symmetry is non-anomalous, and it is this fact that makes
Calabi-Yau target spaces so special.

The simplest examples of Fano varieties are complex projective spaces
$\CC\PP^n$. The corresponding $N=2$ field theories are well studied; in fact,
these models are integrable, in the sense that the exact S-matrix is known~\cite{CPn,CPntwo}.
These theories have only massive vacua.
A more general set of examples is given by Grassmann varieties $G(n,k)$,
which are defined as spaces of complex $k$-planes in an $n$-dimensional complex
vector space. The corresponding $N=2$ field theories are also integrable~\cite{Grassmann}.

An $N=2$ field theory which has a conserved vector (resp. axial) R-current admits a topological A-twist
(resp. B-twist), which yields a 2d topological field theory called the A-model (resp. B-model).
$N=2$ superconformal field theories have both axial and vector R-symmetries, and therefore admit
both kinds of twisting. A-branes and B-branes are ``defined'' as boundary conditions 
which are consistent with A-twist and B-twist, respectively.\footnote{We put
the word ``defined'' in quotes because there is no generally accepted
definition of a boundary condition for a 2d field theory.}
These two sets of branes have
the structure of a category.\footnote{The set of all D-branes is not a category in any natural sense.
The reason is the presence of singular terms in the boundary operator product expansion.}
The space of morphisms is defined as the state
space of topological open strings stretched between pairs of branes. Composition of 
morphisms is defined by means of 3-point correlators in topological open
string theory. Since state spaces of open strings are graded vector
spaces, brane categories are graded categories. In the case of A-branes, 
spaces of morphisms are graded
by the axial R-charge; in the case of B-branes, by the vector R-charge.
For Fano varieties, the axial R-symmetry is generically $\ZZ_2$, so spaces of morphisms
in the category of A-branes are $\ZZ_2$-graded vector spaces. In the Calabi-Yau case, 
the full axial $U(1)$ R-symmetry is non-anomalous, and therefore the category of A-branes is
$\ZZ$-graded. This is the reason one has to work with $\ZZ$-graded Lagrangian submanifolds in the
Calabi-Yau case~\cite{K1}. In the Fano case, one only needs to require that Lagrangian submanifolds
be oriented.

In the Calabi-Yau case, we can also consider the category of B-branes. Since the vector R-current
is non-anomalous, this category is $\ZZ$-graded. In the Fano case, the B-twist is not defined,
and there is no obvious way to define the category of B-branes.

Given a graded category, one can enlarge it by adding for any object $Y$
its shifts $Y[i]$, where 
$i\in\ZZ$ or $i\in \ZZ/2\ZZ$, and defining morphisms as follows:
$$
Mor^k\left(Y_1[i],Y_2[j]\right)=Mor^{k+j-i}(Y_1,Y_2).
$$
In string theory, the R-charge of strings connecting two different branes
is defined only up to an integer constant; changing this constant by $k$
shifts the degree of all morphisms by $k$. The effect of this arbitrariness
is that for any brane $Y$ its shifts $Y[i]$ are automatically included.
This implies that no information is lost if we replace groups of
morphisms with their degree-0 components. This is what one usually does
when working with categories of complexes, such as the derived category.
Nevertheless, in this paper we will keep morphisms of all degrees, since
this conforms better to physical conventions. From this viewpoint, the mathematical
counterpart of the category of B-branes on a Calabi-Yau $X$ is not
$D^b(Coh(X))$, but a $\ZZ$-graded category which is called the completion of $D^b(Coh(X))$
with respect to the shift functor.

For a sigma-model on a Calabi-Yau manifold which is a complete intersection in a toric variety, 
the mirror theory is again a sigma-model of the same kind. For Fano varieties which
are complete intersections in a toric variety,
the mirror theory is a LG model whose target is a non-compact Calabi-Yau~\cite{HV}. 
A general definition of a LG model 
involves, besides a choice of a target manifold, a choice of a holomorphic
function $W$ on this manifold (the superpotential). Thus non-trivial LG
models require non-compact target spaces. This non-compactness usually does
not cause trouble: the important thing is for the critical set of $W$ to
be compact. In general, superpotential breaks vector R-symmetry down
to $\ZZ_2$. Thus the A-twist is not defined, in general. On the other hand, since the canonical bundle
of the target manifold is trivial, the axial R-symmetry is not anomalous,
and the B-twist is well-defined. We expect that the category of A-branes on a Fano variety
is equivalent to the category of B-branes on the mirror LG model.

For example, the mirror of $\CC\PP^n$ is a LG model whose target is
$\left(\CC^*\right)^n$ with the superpotential Eq.~(\ref{WCPn}). This 
superpotential has $n+1$ non-degenerate critical points given by
$$
x_1=x_2=\ldots=x_n=e^{2\pi i k/(n+1)}, \quad k=0,\ldots, n.
$$
The physical interpretation is that the theory has $n+1$ massive vacua.
This agrees with the count of vacua in the $\CC\PP^n$ model. 
Furthermore, the superpotential breaks vector $U(1)_R$ symmetry down to
$\ZZ_2$ given by
$$
\theta_+\ra -\theta_+,\quad \theta_-\ra -\theta_-,
$$  
where $\theta_\pm$ are the usual odd coordinates on the chiral $(2,2)$ superspace.
As a consequence, spaces of morphisms in the category of B-branes in this
LG model are $\ZZ_2$-graded. This is mirror to the fact that morphisms 
in the category of A-branes on $\CC\PP^n$ are $\ZZ_2$-graded.\footnote{In this LG model,
there is in fact an unbroken $\ZZ_{2(n+1)}$ R-symmetry; the $\ZZ_2$ symmetry discussed in the text
is its subgroup. This is mirror to the fact that the $\CC\PP^n$ sigma-model has
non-anomalous axial $\ZZ_{2(n+1)}$ R-symmetry. In this paper we will only keep track
of $\ZZ_2$-gradings.}

As a rule, it is easier to understand B-branes, rather than A-branes. 
Therefore, we now turn to the study of B-branes in LG models, in
the hope that it will illuminate the properties of A-branes on
Fano varieties.

\section{General properties of B-branes in LG models}\label{sec:B}

The classical geometry of B-branes was described in Refs.~\cite{GJS,HIV} (see also Ref.~\cite{ZL}). In
this section we summarize the results of Refs.~\cite{GJS,HIV} which are relevant
for us and discuss some simple consequences.

Let $X$ be the target space of a LG model. On general grounds, it must be
K\"ahler manifold (possibly non-compact). Let the $W$ be a fixed holomorphic
function on $X$ (the superpotential). Let $Y$ be a submanifold of $X$,
and let $E$ be a Hermitian vector bundle over $Y$ with a unitary connection $\nabla$. The rank 
of $E$ will be called the multiplicity of the corresponding D-brane. It is 
shown in Ref.~\cite{HIV} that the triple $(Y,E,\nabla)$ defines a 
classical B-type boundary condition if and only if $Y$ is a complex 
submanifold of $X$, $W$ is constant on $Y$,
and the pair $(E,\bpartial)$, where $\bpartial$ is the anti-holomorphic part 
of $\nabla$, is a holomorphic vector bundle. For example, a point on
$X$ together with a choice of multiplicity $r\in\NN$ defines a B-type boundary 
condition.

The class of B-branes described in the previous paragraph does not exhaust
all possible B-branes. But it appears plausible that all B-branes
can be obtained as bound states of the branes described above. 

It was noticed in Ref.~\cite{H} (see also Ref.~\cite{HICM}) that most of the ``classical'' B-branes
should be regarded as zero objects in the category of B-branes. A classical B-brane
is isomorphic to the zero object if and only if the space of its endomorphisms is 
zero-dimensional, i.e. when there are no supersymmetric open string states
connecting the brane with itself. In this case one says that world-sheet supersymmetry
is spontaneously broken. For example, it is
explained in Refs.~\cite{Hbook,H,HICM} that if $Y$ is a point on $X$ which is
not a critical point of $W$, then B-type supersymmetry is spontaneously
broken, and therefore $Y$ is isomorphic to the zero object in the category of 
B-branes.

This phenomenon reduces enormously the number of B-branes that one needs
to consider, and makes it plausible that the whole category can be
described combinatorially, using only the number and type of critical points
of $W$. To substantiate this claim, we first notice that to any B-brane
in the class described above one can assign a complex number, the value
of $W$ on this brane. Further, there are no non-zero morphisms between
branes with different values of $W$, because any string connecting such
branes will have non-zero energy and will not be supersymmetric~\cite{Hbook,HICM}.
(Unlike in the case of A-branes, there is no central charge in the 
supersymmetry algebra, and supersymmetric states must have zero energy.)
Thus the category of B-branes can be regarded as a family of categories
parametrized by $\CC$, and the categories at different points in $\CC$ do
not ``talk'' to each other. 

Second, zero-energy classical
configurations of an open string must be constant maps from the interval to $Y$, such
that the potential energy $|\partial W|^2$ vanishes. This implies
that unless a B-brane passes through a critical point of $W$, there are
no supersymmetric states for strings connecting this B-brane to any other
B-brane (including itself). It follows that categories corresponding to 
non-critical values of $W$ are trivial (contain only the zero object).

Now let us assume that all critical points of $W$ are isolated. By scaling
up the K\"ahler form, we can make the semi-classical approximation arbitrarily
good. This means that wave-functions of all string states will be arbitrarily
well localized near a particular critical point of $W$, and the overlap
between wave-functions associated to different critical points will be
arbitrarily small. Since topological correlators do not depend on the
K\"ahler form, it is clear that morphisms between B-branes can be computed using only the
leading terms in the Taylor expansion of $W$ around the critical points.\footnote{In fact, in
Refs.~\cite{HV,H,HICM} there is a proposal how to compute spaces of morphisms between B-branes using
a deformation of the Dolbeault complex by $\partial W$.}

To be more precise, we can attach a category to each isolated critical point of $W$ as follows:
we replace $W$ by a polynomial which has the same singularity, and consider the category of
B-branes on an affine space with this polynomial superpotential. Now let us form the direct sum of
such categories over all critical points of $W$ and call it ${\mathsf C}_{tot}$. There is an obvious map
which associates to any B-brane an object of $\C_{tot}$. Invariance of topological correlators
under variations of the K\"ahler form means that this map extends to a functor, and this functor
is full and faithful. In other words, the category of B-branes is a full sub-category of $\C_{tot}$.

In particular, when all critical points of $W$ are non-degenerate (i.e. when
all vacua are massive), the problem reduces to understanding
B-branes in the LG model with target $\CC^n$ and superpotential 
\begin{equation}\label{Wfreen}
W=z_1^2+\ldots+ z_n^2.
\end{equation}
The corresponding bulk theory is free, but since the boundary conditions
need not be linear, the problem of determining all B-branes is far from trivial. In this paper
we will study B-branes which correspond to linear boundary conditions. Some such branes
have been considered in Refs.~\cite{H,Hbook,HICM}.
We will see below that if Kontsevich's conjecture is true, then these
branes generate the whole category of B-branes. 

Note that the LG superpotential Eq.~(\ref{WCPn}) satisfies the conditions
stated above. Thus, assuming mirror symmetry,
we can gain information about A-branes on $\CC\PP^n$ by studying B-branes
in the free LG model with the superpotential Eq.~(\ref{Wfreen}).
In the case $n=1$ this has been done in Ref.~\cite{H}; in that case
the category of A-branes is independently known, and one can see that
the mirror conjecture holds true. In Section~\ref{sec:F}
we discuss the less trivial case $n=2$.

Before continuing, let us make some further comments on the relation between $\C_{tot}$
and the category of B-branes. We do not claim that the two categories are equivalent, only that
the latter is a full sub-category of the former. This means that each B-brane can be regarded
as a direct sum of ``local'' B-branes attached to critical points, but not every direct sum
is a valid B-brane. We will see in Section~\ref{sec:F} some examples where the category of B-branes
is strictly smaller than $\C_{tot}$. Nevertheless, since each B-brane behaves as a composite of
``local'' B-branes, it is reasonable to enlarge the category of B-branes by allowing arbitrary
sums of ``local'' B-branes. Then the category of B-branes becomes equivalent to $\C_{tot}$.
{}From a purely algebraic standpoint, this is a very natural procedure (c.f. a discussion in Ref.~\cite{Laz} concerning
reducible and irreducible branes), but the drawback is that the new branes lack a clear
geometric interpretation. Section~\ref{sec:F} contains a further discussion of this issue.

\section{B-Branes in the LG model with $W=xy$}
\label{sec:C}

\subsection{Preliminaries}

We begin by recalling the results of Ref.~\cite{H} concerning B-branes in the
simplest LG model with the superpotential $W=z^2$. In this case, the only allowed
B-branes are D0-branes located at $z=0$. It has been shown in Ref.~\cite{H} 
that the space of endomorphisms of a single D0-brane is two-dimensional, with 
one-dimensional even subspace and one-dimensional odd subspace. As a graded 
algebra, it is generated over $\CC$ by the identity and an odd element 
$\theta$ with the relation
$$
\theta^2=1.
$$
This is a Clifford algebra $Cl(1,\CC)$. If we take $N$ D0-branes, then
the algebra of endomorphisms becomes
$$
Cl(1,\CC)\otimes Mat(N,\CC),
$$
where $Mat(N,\CC)$ is the algebra of $N\times N$ complex matrices.

We are interested in the next simplest LG model with the superpotential
$W=xy$. Again, D0-branes must be localized at $x=y=0$, and it has been shown
in Ref.~\cite{H} that the algebra of endomorphisms of a single D0-brane
is generated by two odd elements $\theta_1,\theta_2$ with
the relations
$$
\theta_1\theta_2+\theta_2\theta_1=1,\quad \left(\theta_1\right)^2=
\left(\theta_2\right)^2=0.
$$
This is a Clifford algebra $Cl(2,\CC)$. More invariantly, if we denote
by $V$ the complex vector space which is the target space of our LG model,
we can say that fermion zero modes for the open string take values in $V$. The Hessian
of the superpotential
$$
Q_{ij}=\partial_i\partial_j W(0)
$$
defines a non-degenerate symmetric bilinear form on $V$, and the
endomorphism algebra of the D0-brane is the Clifford algebra associated
to the pair $(V,Q)$. (Some standard facts about Clifford algebras and their
modules are described in the Appendix. We will freely use these facts in what follows.)

But in this case there can also be B-branes of higher dimension,
namely D2-branes. (This is briefly discussed in Ref.~\cite{HICM}.)
Irreducible D2-branes are irreducible components of the critical level 
set $W=0$, which is a singular quadric
$$
xy=0.
$$
Thus there are two candidate irreducible D2-branes, given by $x=0$ and $y=0$, 
respectively.
Our immediate goal is to compute their endomorphisms, as well as morphisms between 
D2-branes and D0-branes. In physical terms, we will compute the spectrum
and disk correlators of the topological open string with appropriate boundary
conditions.

\subsection{Equations of motion and SUSY transformations}

We consider the \LG model on $\RR\times[0,\pi]$ with two chiral superfields $\Phi_1$ and $\Phi_2$. 
The superpotential assumes the following form 
$$W(\Phi) = 2m\Phi_1\Phi_2.$$
We include a positive factor $2m$ in the superpotential in order to keep track of dimensions of topological correlators later. 
In physical terms, $m$ is a measure of the mass gap in the \LG model.

Assuming the standard K\"ahler potential $K=|\Phi_1|^2+|\Phi_2|^2,$ the world-sheet action reads
\begin{eqnarray}
	 S &=& \frac1{2\pi}\int_{\RR\times[0,\pi]}\!\!\d^2x \l\{\sum_{\a=1}^{2}\l(|\partial_t\phi^{\a}|^2-
|\partial_\sm\phi^\a|^2+i\psi_-^{\bar{\a}}\partial_+\psi_-^\a +i\psi_+^{\bar{\a}}\partial_-\psi_+^\a\r)\r. \nonumber\\
	&& \l.-\,|m\phi^1|^2 -|m\phi^2|^2- m\l(\psi_+^1\psi_-^2+\psi_+^2\psi_-^1\r) -
\bar{m}\l(\psi_-^{\bar1}\psi_+^{\bar2}+\psi_-^{\bar2}\psi_+^{\bar1}\r)\r\},\no
\end{eqnarray}
where $\phi^\a$ are the bosonic components of $\Phi_\a$, and $\psi^\a$ are their fermionic partners. 
The world-sheet parametrization $(t,\sm)$ is such that $t$ is the world-sheet time. 

{}From the bosonic Lagrangian density  
$$\L_B = \sum_{\a=1,2}\l(|\dot\phi_\a|^2-|\phi_\a'|^2-m^2|\phi_\a|^2\r)$$
one readily obtains the EOM's for $\phi$
$$\ddot{\phi}_\a-\phi_\a''+m^2\phi_\a=0, \qquad \quad \a=1,2.$$
Similarly, in terms of new variables 
$$b_\a = \frac{\psi_-^{\a}+\psi_+^{\a}}{\sqrt2},\qquad\quad c_\a = \frac{\psi_-^{\a}-\psi_+^{\a}}{\sqrt2}, \qquad \a=1,2,$$
the fermionic Lagrangian density can be written as
\begin{eqnarray}
	\L_F &=& i\bar{b}_1\dot{b}_1+i\bar{b}_2\dot{b}_2+i\bar{c}_1\dot{c}_1+i\bar{c}_2\dot{c}_2\nonumber\\
	&& + \bar{b}_1(i\partial_\sm c_1 +\bar{m}\bar{c_2}) + (-i\partial_\sm\bar{c}_1 + mc_2)b_1\nonumber\\
	&& + \bar{b}_2(i\partial_\sm c_2 +\bar{m}\bar{c_1}) + (-i\partial_\sm\bar{c}_2 + mc_1)b_2\nonumber
\end{eqnarray}

The fermionic EOM's are given by
\begin{equation}
	\begin{array}{l}
	i\dot{b}_1 + ic_1' + m\bar{c}_2 = 0,\\
	i\dot{b}_2 + ic_2' + m\bar{c}_1 = 0,\\
	i\dot{c}_1 + ib_1' - m\bar{b}_2 = 0,\\
	i\dot{c}_2 + ib_2' - m\bar{b}_1 = 0.\end{array}
	\label{eq:feom}
\end{equation}

B-type supersymmetry transformations are well known (see, for example, Ref.~\cite{HIV}) and look as follows:
\begin{equation}
	\begin{array}{l}\delta\phi_1 = \sqrt2\ep b_1, \quad \delta b_1 = -\sqrt2\,i\bar{\ep}\dot{\phi}_1, 
\quad \delta c_1 = \sqrt2\,i\bar{\ep}\phi_1' + \sqrt2\ep\bar{m}\bar{\phi}_2,\\ 
\delta\phi_2 = \sqrt2\ep b_2, \quad \delta b_2 = -\sqrt2\,i\bar{\ep}\dot{\phi}_2, \quad \delta c_2 = \sqrt2\,i\bar{\ep}\phi_2' + 
\sqrt2\ep\bar{m}\bar{\phi}_1.\end{array}
	\label{eq:susy}
\end{equation}

Now we will perform canonical quantization of this system with various
boundary conditions which correspond to D0-D0 strings, D2-D2 strings, and
D0-D2 strings.

\subsection{Spectrum of D0-D0 strings}
We would like to find supersymmetric states of D0-D0 strings, since these
correspond to endomorphisms of the D0-brane.
The relevant boundary conditions are
$$
\l\{\begin{array}{l}\phi_1=\phi_2=0\\b_1=b_2=0\end{array}\r.\qquad\quad 
\sm=0,\,\pi.
$$

\vt
First consider the bosonic degrees of freedom. The boundary conditions give 
the following mode expansions for $\phi$ and its conjugate momentum 
$\pi_\phi$:
\begin{eqnarray}
	\phi_\a &=& \frac1{\sqrt{\pi}}\sum_{n=1}^{\infty} 
	\frac{i}{\sqrt{\om_n}}\l(a_{\a,n}-\ta_{\a,n}^\dagger\r)
	\sin n\sm\nonumber\\
	\pi_{\phi_\a} &=& \partial_t\bar{\phi} = 
	\frac1{\sqrt\pi}\sum_{n=1}^{\infty} \sqrt{\om_n}
	\l(\ta_{\a,n}+a_{\a,n}^\dagger\r)\sin n\sm\nonumber
\end{eqnarray}
where 
$$\om_n= \sqrt{n^2 + m^2}.$$

To quantize, we impose the following commutation relations:
\begin{equation}
	\label{eq:comm_a}
	\Big[a_{\a,n}, a_{\b,m}^\dagger\Big]=\Big[\ta_{\a,n}, 
	\ta_{\b,m}^\dagger\Big]=\delta_{\a\b}\delta_{mn},
\end{equation}
with all other commutators vanishing. It is easy to check that (\ref{eq:comm_a}) is compatible with the following 
canonical commutation relations 
\begin{equation}
	[\phi_\a(\sm), \pi_{\phi_\b}(\sm')] = i\delta_{\a\b}\cdot 
	\frac2{\pi}\sum_{n=1}^{\infty} \sin n\sm\sin n\sm'
	= i\,\delta_{\a\b}\,\delta(\sm-\sm')
	\label{eq:can_phi}
\end{equation}

The bosonic Hamiltonian is given by
\begin{eqnarray}
	H_B &=& \int\d\sm \sum_{\a=1,2}\l(|\dot\phi_\a|^2+|\phi_\a'|^2+
	m^2|\phi_\a|^2\r)\nonumber\\
	&=& \sum_{n=1}^{\infty} \om_n\l(a_{1,n}^\dagger a_{1,n} + 
	\ta_{1,n}^\dagger\ta_{1,n}+a_{2,n}^\dagger a_{2,n} + 
	\ta_{2,n}^\dagger\ta_{2,n} + 2\r)\nonumber
\end{eqnarray}
where the additive constant ``$+2$'' in the sum is the bosonic zero point 
energy. We shall see later that it is exactly canceled by the fermionic 
zero point energy. 

Next we consider the fermionic degrees of freedom.
It will turn out convenient to use the following combinations as new 
dynamical variables:
$$b_{\pm} = \frac{b_1\pm i\bar{b}_2}{\sqrt2}, \qquad\quad c_{\pm} = 
\frac{c_1\pm i\bar{c}_2}{\sqrt2}.$$
The main advantage of using $b_\pm$ and $c_\pm$ is that the EOM's 
for the unbarred quantities are decoupled from those for the barred quantities. 
The mode expansions for these fields have the following form:
\begin{eqnarray}
	b_+ &=& \frac{i}{\sqrt\pi}\sum_{n=1}^\infty \frac{n+im}{\om}\l(\a_{2,n}
	-\taa_{2,n}^\dagger\r)\sin n\sm,\no\\
	b_- &=& \frac{i}{\sqrt\pi}\sum_{n=1}^\infty \frac{n-im}{\om}\l(\a_{1,n}
	-\taa_{1,n}^\dagger\r)\sin n\sm,\no\\
	c_+ &=& \frac1{\sqrt\pi}\sum_{n=1}^\infty \frac{1}{n-im}(n\cos n\sm
	+m\sin n\sm)\l(\a_{2,n}+\taa_{2,n}^\dagger\r) + \la_+ e^{m\sm},\no\\
	c_- &=& \frac1{\sqrt\pi}\sum_{n=1}^\infty \frac{1}{n+im}(n\cos n\sm
	-m\sin n\sm)\l(\a_{1,n}+\taa_{1,n}^\dagger\r) + \la_- e^{-m\sm}.\no
\end{eqnarray}
To fix the commutation rules for the oscillators we impose the canonical 
commutation relations for $b$:
\begin{gather}
\{b_+,b_+\}=\{b_-,b_-\}=\{b_+,b_-\}=\{b_+,\bar{b}_-\}=\{b_-,\bar{b}_+\}=0\no\\	
\{b_+(\sm),\bar{b}_+(\sm')\} = \{b_-(\sm),\bar{b}_-(\sm')\} = \delta(\sm-\sm').
 \label{eq:can_b}
\end{gather}
A convenient choice of compatible commutation rules for oscillators is
\begin{equation}
\big\{\a_{i,n},\a^\dagger_{i',n'}\big\} = 
\big\{\taa_{i,n},\taa^\dagger_{i',n'}\big\} = \delta_{i\,i'}\delta_{nn'},
	\label{eq:comm_alpha}
\end{equation}
with all others vanishing. One can easily check that the canonical commutation
relations for $c$ fields are also respected, provided that the following relations are imposed:
\begin{gather}
\big\{\zeta,\zeta\big\}=\big\{\eta,\eta\big\}=\big\{\zeta,\eta\big\} = \big\{\zeta,\eta^\dagger\big\}=0,\no\\
\big\{\zeta,\zeta^\dagger\big\} = \big\{\eta,\eta^\dagger\big\} = 1.\no
\end{gather}
where $\eta$ and $\zeta$ are defined by
$$
\la_+ \equiv \zeta \sqrt{\frac{m}{\sinh m\pi}}e^{-\pi m/2}, \qquad 
\la_- \equiv \eta \sqrt{\frac{m}{\sinh m\pi}}e^{\pi m/2}.
$$
The fermionic Hamiltonian is
\begin{align}
	H_F &= \int_0^\pi\!\d\sm \, \big[\l(ic_1'+m\bar{c}_2\r)\bar{b}_1 
	+\l(ic_2'+m\bar{c}_1\r)\bar{b}_2 + {\rm h.c.} \big]\no\\
	&=  \int_0^\pi\!\d\sm \, \l[\, i(c_+'-mc_+)\bar{b}_+ 
	+ i(c_-'+mc_-)\bar{b}_- +  {\rm h.c.} \r]\no\\
	&= \sum_{n=1}^\infty \; \om_n \l(\a^\dagger_{1,n}\a_{1,n}+
	\a^\dagger_{2,n}\a_{2,n}+\taa^\dagger_{1,n}\taa_{1,n}+\taa^\dagger_{2,n}\taa_{2,n}-2\r).\no
\end{align}
Note that the additive constant $-2$ cancels the bosonic zero 
point energy. The Hamiltonian is diagonalized in the Fock basis, 
and the zero-energy states are 
$$|0\rangle, \quad \zeta^\dagger|0\rangle ,\quad \eta^\dagger|0\rangle, \quad 
\eta^\dagger\zeta^\dagger|0\rangle.$$
The supercharge $Q$  can also be expanded in terms of oscillators. It can be
shown that each term in the expansion contains an
annihilation operator for non-zero modes, and that $Q$ does not depend on 
zero-mode oscillators $\zeta$ and $\eta$. Therefore $Q$ annihilates 
all four ground states.

\subsection{Spectrum of D2-D2 strings}

Since we have two different D2-branes related by a symmetry, there are two 
inequivalent possibilities:
either our string begins and ends on the same D2-brane, or it begins on one
D2-brane, and ends on the other D2-brane. The first situation corresponds
to endomorphisms of a D2-brane, while the second one corresponds to morphisms
from one D2-brane to the other one.

First we consider the case when both boundaries end on the same brane, say,
the one given by the equation $\Phi_1=0$. 
The relevant boundary conditions are
$$
\Big\{\begin{array}{l}\phi_1=0\\b_1=0\end{array}\qquad 
{\rm and} \qquad \Big\{\begin{array}{l}\partial_\sm\phi_2=0\\
c_2=0\end{array}\qquad {\rm at} \;\;\; \sm=0,\,\pi.
$$

First let us look at the bosons. The mode expansions for $\phi_1$ and its
conjugate momentum are the same as before, while for $\phi_2$ and its conjugate
momentum they are given by
\begin{align}
\phi_2 &= \frac1{\sqrt{2\pi m}}\l(a_{2,0}+\ta_{2,0}^\dagger\r)+
\sum_{n=1}^{\infty} \frac{1}{\sqrt{\pi\om_n}}\l(a_{2,n}+
\ta_{2,n}^\dagger\r)\cos n\sm,\nonumber\\
\pi_{\phi_2}&= i\sqrt{\frac{m}{2\pi}}\l(a_{2,0}^\dagger-\ta_{2,0}\r) + i\sum_{n=1}^{\infty} 
\sqrt{\frac{\om_n}{\pi}}\l(a_{2,n}^\dagger- \ta_{2,n}\r)\cos n\sm\nonumber
\end{align}
Canonical commutation relations for bosonic fields imply the
commutation relations Eq.~(\ref{eq:comm_a}) for the oscillators. 
In terms of oscillators the bosonic Hamiltonian is
\begin{eqnarray}
	H_B &=& \sum_{n=1}^\infty 
	\om_n \l(a_{1,n}^\dagger a_{1,n}+\ta_{1,n}^\dagger\ta_{1,n}+1\r)\nonumber\\
	&+& \sum_{n=0}^\infty 
	\om_n \l(a_{2,n}^\dagger a_{2,n}+\ta_{2,n}^\dagger\ta_{2,n}+1\r).\nonumber
\end{eqnarray}

For the fermions, the mode expansions are given by
\begin{eqnarray}
	b_1 &=& \frac{i}{\sqrt{2\pi}}\sum_{n=1}^\infty 
	\l[\frac{n+im}{\om_n}\l(\a_{2,n}-
	\taa_{2,n}^\dagger\r) + \frac{n-im}{\om_n}\l(\a_{1,n}-
	\taa_{1,n}^\dagger\r)\r]\sin n\sm,\no\\
	b_2 &=& \frac{-1}{\sqrt{2\pi}} \l( \taa_{0} - \a_0^\dagger\r)\no\\
	&+& \frac{-i}{\sqrt{2\pi}}\sum_{n=1}^\infty 
	\l[-\frac{n+im}{\om_n}\l(\taa_{1,n}-\a_{1,n}^\dagger\r) + 
	\frac{n-im}{\om_n}\l(\taa_{2,n}-\a_{2,n}^\dagger\r)\r]\cos n\sm,\no\\
	c_1 &=& \frac{1}{\sqrt{2\pi}}\l(\a_{0}+\taa_{0}^\dagger\r) + \frac{1}{\sqrt{2\pi}}\sum_{n=1}^\infty \l(\a_{1,n}+
\a_{2,n}+\taa_{1,n}^\dagger+\taa_{2,n}^\dagger\r)\cos n\sm,\no\\
	c_2 &=&  \frac{1}{\sqrt{2\pi}}\sum_{n=1}^\infty 
	\l(\taa_{2,n}-\taa_{1,n}+\a_{2,n}^\dagger-\a_{1,n}^\dagger\r)\sin n\sm .\no
\end{eqnarray}
The canonical commutation relations for the fields $b_i$ and $c_i$ are
equivalent to the following commutation relations for the oscillators:
\begin{gather}
	\big\{\a_{i,n},\a^\dagger_{i',n'}\big\} = 
	\big\{\taa_{i,n},\taa^\dagger_{i',n'}\big\} = 
	\delta_{i\,i'}\delta_{nn'}, \qquad n=1,2,\ldots,\no\\
	\big\{\a_0,\a_0^\dagger\big\} =
	\big\{\taa_0,\taa_0^\dagger\big\} = 1,\no
\end{gather}
with all others vanishing. 
The fermionic Hamiltonian can be shown to be
\begin{eqnarray}
	H_F &=& \int_0^\pi\!\d\sm \, \big[\l(ic_1'+m\bar{c}_2\r)\bar{b}_1 
	+\l(ic_2'+m\bar{c}_1\r)\bar{b}_2 + {\rm h.c.} \big]\no\\
	&=& \sum_{n=1}^\infty \; \om_n \l(\a^\dagger_{1,n}\a_{1,n}+
	\a^\dagger_{2,n}\a_{2,n}+\taa^\dagger_{1,n}\taa_{1,n}+
	\taa^\dagger_{2,n}\taa_{2,n}-2\r)\nonumber\\
	&& +\; m\l(\a_0^\dagger\a_0 + \taa_0^\dagger\taa_0-1\r).\no
\end{eqnarray}
The fermionic zero point energy cancels the bosonic zero point energy, and we
see that there is a unique state with zero energy: the Fock vacuum.
For the same reason as in the D0-D0 case, this state is supersymmetric
(is annihilated by the supercharge). 

Now consider the case when one end of the string ($\sm=0$) is attached to 
$\Phi_1=0$, and the other one ($\sm=\pi$) is attached to $\Phi_2=0$. 
The boundary conditions are
\begin{equation}
	\l\{\begin{array}{l}\phi_1(0)=\partial_\sm\phi_1(\pi)=0\\
		b_1(0)= c_1(\pi)=0\end{array}\r.
	\qquad{\rm and} \qquad  \l\{\begin{array}{l}\partial_\sm\phi_2(0)=\phi_2(\pi)=0\\
		c_2(0)=b_2(\pi)=0\end{array}\r.\nonumber
\end{equation}
The mode expansions for the bosons are
\begin{eqnarray}
	\phi_1 &=& \sum_{n=1}^{\infty} \frac{i}{\sqrt{\pi\om_n}}\,\l(a_{1,n}-
	\ta_{1,n}^\dagger\r)\sin k_n\sm,\no\\
	\pi_{\phi_1} &=& \sum_{n=1}^{\infty} \sqrt{\frac{\om_n}{\pi}}\,\l(\ta_{1,n}+ a_{1,n}^\dagger\r)\sin k_n\sm,\no\\
	\phi_2 &=& \sum_{n=1}^{\infty} \frac1{\sqrt{\pi\om_n}}\l(a_{2,n}+
	\ta_{2,n}^\dagger\r)\cos k_n\sm,\nonumber\\
	\pi_{\phi_2} &=& \sum_{n=1}^{\infty} i\sqrt{\frac{\om_n}{\pi}}\,\l(-\ta_{1,n}+ a_{1,n}^\dagger\r)\cos k_n\sm,\no
\end{eqnarray}
while for the fermions they are
\begin{eqnarray}
	b_1 &=& \frac{i}{\sqrt{2\pi}}\sum_{n=1}^\infty 
	\l[\frac{k_n+im}{\om_n}\l(\a_{2,n}-
	\taa_{2,n}^\dagger\r) + \frac{k_n-im}{\om_n}\l(\a_{1,n}-
	\taa_{1,n}^\dagger\r)\r]\sin k_n\sm,\no\\
	b_2 &=& \frac{1}{\sqrt{2\pi}}\sum_{n=1}^\infty 
	\l[\frac{k_n+im}{\om_n}\l(\taa_{1,n}-\a_{1,n}^\dagger\r) - 
	\frac{k_n-im}{\om_n}\l(\taa_{2,n}-\a_{2,n}^\dagger\r)\r]\cos k_n\sm,\no\\
	c_1 &=& \frac{1}{\sqrt{2\pi}}\sum_{n=1}^\infty \l(\a_{1,n}+\a_{2,n}+\taa_{1,n}^\dagger+\taa_{2,n}^\dagger\r)\cos k_n\sm,\no\\
	c_2 &=&  \frac{i}{\sqrt{2\pi}}\sum_{n=1}^\infty 
	\l(\taa_{1,n}-\taa_{2,n}+\a_{1,n}^\dagger-\a_{2,n}^\dagger\r)\sin k_n\sm,\no
\end{eqnarray}
where 
$$k_n = n-1/2, \qquad \om_n = \sqrt{k_n^2+m^2}.$$
One can show as before that commutation relations (\ref{eq:comm_a}) and (\ref{eq:comm_alpha}) yield all the canonical commutation relations,
and the total Hamiltonian is diagonalized in the Fock basis as follows:
$$H = \sum_{i=1}^2 \sum_{n=0}^\infty 
	\om_n \l(a_{i,n}^\dagger a_{i,n}+\ta_{i,n}^\dagger\ta_{i,n}+\a_{i,n}^\dagger \a_{i,n}+\taa_{i,n}^\dagger\taa_{i,n}\r).$$
Again there is a single ground state which is annihilated by the supercharge.

\subsection{Spectrum of D0-D2 strings} 
\label{sec:d0d2}

The boundary conditions for the bosons are
\begin{equation}
	\phi_1(0)=\phi_1(\pi)=0,\qquad 
	   \phi_2(0)=\partial_\sm\phi_2(\pi)=0.
\end{equation}
The corresponding mode expansions are
\begin{eqnarray}
	\phi_1 &=& \sum_{n=1}^{\infty} \frac{i}{\sqrt{\pi\om_{1,n}}}\,
	\l(a_{1,n}-\ta_{1,n}^\dagger\r)\sin n\sm,\nonumber\\
	\pi_{\phi_1} &=& \sum_{n=1}^{\infty} \sqrt{\frac{\om_{1,n}}{\pi}}\,
	\l(\ta_{1,n}+a_{1,n}^\dagger\r)\sin n\sm,\no\\
	\phi_2 &=& \sum_{n=1}^{\infty} \frac{i}{\sqrt{\pi\om_{2,n}}}\,
	\l(a_{2,n}-\ta_{2,n}^\dagger\r)\sin (n-1/2)\sm,\nonumber\\
	\pi_{\phi_2} &=& \sum_{n=1}^{\infty} \sqrt{\frac{\om_{2,n}}{\pi}}\,
	\l(\ta_{2,n}+a_{2,n}^\dagger\r)\sin (n-1/2)\sm,\no
\end{eqnarray}
where
$$\om_{1,n} = \sqrt{n^2+m^2}, \qquad \om_{2,n} = \sqrt{(n-1/2)^2+m^2}.$$
Imposing (\ref{eq:can_phi}), we infer that the bosonic oscillators 
obey (\ref{eq:comm_a}). The bosonic Hamiltonian is given by
\begin{eqnarray}
	H_B &=& \sum_{n=1}^\infty \om_{1,n} \l(a_{1,n}^\dagger a_{1,n}+
	\ta_{1,n}^\dagger\ta_{1,n} + 1\r)\no\\
	&+& \sum_{n=1}^\infty \om_{2,n} \l(a_{2,n}^\dagger a_{2,n}+
	\ta_{2,n}^\dagger\ta_{2,n} + 1\r).\no
\end{eqnarray}

Now let us consider fermions. The boundary conditions imposed by supersymmetry 
are 
$$b_1(0)=b_1(\pi)=b_2(0)=c_2(\pi)=0,$$
which, when combined with the EOM's, give the following boundary 
conditions for $b$ and $c$ fields separately:
\begin{equation}
	b_1(0)=b_1(\pi)=0, \qquad b_2(0)=b_2'(\pi) = 0,
\end{equation}
\begin{equation}
	\l\{\begin{array}{l}c_1'(0)-im\bar{c}_2(0) = 
	c_2'(0)-im\bar{c}_1(0) = 0,\\
	c_1'(\pi) = c_2(\pi) = 0.\end{array}\r.
	\label{eq:b.c._c}
\end{equation}
The mode expansions are
\begin{eqnarray}
	b_1 &=& \frac{i}{\sqrt{\pi}}\sum_{n=1}^\infty \frac{k_{1,n}-im}{\om_{1,n}}\l(\a_n-\taa_n^\dagger\r)\sin k_{1,n}\sm,\no\\
	b_2 &=& \frac{1}{\sqrt{\pi}}\sum_{n=1}^\infty \frac{k_{2,n}+im}{\om_{2,n}}\l(\tbb_n-\b_n^\dagger\r)\sin k_{2,n}\sm,\no\\
	c_1 &=& \frac1{\sqrt{\pi}}\sum_{n=1}^\infty \frac{k_{1,n}}{k_{1,n}+im}\l(\a_n+\taa_n^\dagger\r)\cos k_{1,n}\sm\no\\
	&-& \frac1{\sqrt{\pi}}\sum_{n=1}^\infty \frac{m}{k_{2,n}+im}\l(\b_n+\tbb_n^\dagger\r)\sin k_{2,n}\sm + \la\cosh m(\pi-\sm),\no\\
	c_2 &=& \frac{i}{\sqrt{\pi}}\sum_{n=1}^\infty \frac{m}{k_{1,n}-im}\l(\taa_n+\a_n^\dagger\r)\sin k_{1,n}\sm\no\\
	&-& \frac{i}{\sqrt{\pi}}\sum_{n=1}^\infty \frac{k_{2,n}}{k_{2,n}-im}\l(\tbb_n+\b_n^\dagger\r)\cos k_{2,n}\sm -i\la^\dagger
\sinh m(\pi-\sm),\no
\end{eqnarray}
where
$$k_{1,n}=n, \qquad k_{2,n}=n-1/2.$$
Imposing canonical commutation relations on the fields implies the following
commutation relations for the oscillators:
\begin{equation}
	\big\{\a_n, \a_{n'}^\dagger\big\} = \big\{\taa_n, 
	\taa_{n'}^\dagger\big\}=\big\{\b_n, \b_{n'}^\dagger\big\} = \big\{\tbb_n, \tbb_{n'}^\dagger\big\} = \delta_{nn'},
\label{eq:comm_d0d2}
\end{equation}
$$ \big\{\la,\la^\dagger\big\}=\frac{m}{\sinh 2m\pi}.$$
All other anti-commutators vanish.

In view of the commutation relations for $\la$ and $\la^\dagger$, we set
$$\la = \sqrt{\frac{m}{\sinh 2m\pi}}\;\zeta,$$
so that 
$$\big\{\zeta,\zeta\big\}=0, \qquad \big\{\zeta,\zeta^\dagger\big\}=1.$$
The fermionic Hamiltonian has the form
\begin{eqnarray}
	H_F &=& \int_0^\pi\!\d\sm \, \big[\l(ic_1'+m\bar{c}_2\r)\bar{b}_1 +\l(ic_2'+m\bar{c}_1\r)\bar{b}_2 + {\rm h.c.} \big]\no\\
	&=& \sum_{n=1}^\infty  \om_{1,n} \l(\a^\dagger_{n}\a_{n}+\taa^\dagger_{n}\taa_{n}-1\r) + \sum_{n=1}^\infty \om_{2,n} 
\l( \b^\dagger_{n}\b_{n} + \tbb^\dagger_{n}\tbb_{n}-1\r).\no
\end{eqnarray}
As before the bosonic zero-point energy is canceled by the fermionic zero-point energy.  There are two zero-energy states:
$$|0\rangle \qquad {\rm and}\qquad \zeta^\dagger|0\rangle.$$
Again it can be shown that they are annihilated by the supercharge. 
Therefore both ground states are supersymmetric.

\section{Topological correlators in the LG model $W=xy$}\label{sec:Cp}

\subsection{Topological B-twist}
The \LG model admits a topological twist to yield the so-called B-model.  This topological twist turns the world-sheet spinor
fields $\psi$ and $\bar{\psi}$ of the original LG model into a pair of sections  of the pullback bundle $\Phi^*(T^{0,1}X)$, 
which we denote by 
$\eta,\theta$,  and a world-sheet one-form $\rho$ with values in $\Phi^*(T^{1,0}X)$. The BRST transformations of the twisted 
fields are
\begin{eqnarray}
	\delta\phi^i &=& 0,\nonumber\\
	\delta\phi^{\bar{i}} &=& \bar{\ep}\eta^{\bar{i}},\nonumber\\
	\delta\eta^{\bar{i}} &=& 0,\nonumber\\
	\delta\theta^{\bar{i}} &=& \bar{\ep}\partial^{\bar{i}}W,\nonumber\\
	\delta\rho^i &=& i\bar{\ep}\d\phi^i.\nonumber
\end{eqnarray}
To make connection with the fields in the original \LG theory, 
we note that $\eta$ and $\theta$ are the twisted versions of $\bar{b}$ and 
$\bar{c}$ respectively, while $\rho$ comes from $b$ and $c$. We also adopt 
the common notation $\th_i=g_{i\bar{j}}\th^{\bar{j}}=\th^{\bar{i}}$.

The local physical observables are in one-to-one correspondence with the 
BRST-cohomology, i.e. local quantities which are BRST invariant but not BRST 
exact. It is easy to see from the above BRST transformations that in the bulk 
the physical observables correspond to holomorphic functions of $\phi$ modulo  
$\d W(v)$, where $v$ is an arbitrary holomorphic vector field. 
There are no additional local observables from the fermionic fields, 
as long as $W$ is nontrivial. In particular, when $X\simeq\CC^n$,
the space of bulk observables is $\CC[x_1,\ldots,x_n]/I$, 
where $I$ is the ideal generated by the first partial derivatives of $W$.
In the boundary sector, where $W$ is 
constrained to be constant, additional observables will arise from 
the $\theta$ fields.

We now specialize to the D0 and D2 branes studied above.

\subsection{Boundary Observables Associated with the D0-brane}

We consider the boundary component which is mapped to the D0 brane located at $\phi_1=\phi_2=0$. 
The boundary conditions require, among other things, that
$$\phi_1=\phi_2=\eta^{\bar1}=\eta^{\bar2}=0 \qquad\quad \mbox{at the boundary}.$$
Therefore boundary observables can only come from the $\theta$ fields. From the BRST transformation
$$\delta\theta_1 = 2\bar{\ep}m\,\phi_2, \qquad \delta\theta_2 = 2\bar{\ep}m\,\phi_1$$
one sees immediately that both $\th_1$ and $\th_2$ are BRST invariant on the boundary. Let us denote the restriction of 
$\th$ to the boundary by the same letter $\th$. Thus the ring of boundary observables is
generated by $\th_1$ and $\th_2$.

\subsection{Boundary Observables Associated with the D2-brane}

Without loss of generality, we may assume that the D2 brane sits
at the locus $\phi_1=0$. The relevant boundary conditions read
$$\phi_1=\partial_\sm\phi_2=\eta^{\bar1}=\th_2=0 \qquad\quad \mbox{at the boundary}.$$
{}From this one sees that $\th_2$ no longer gives rise to a boundary degree of 
freedom. Also, since $\phi_2$ is not constrained to vanish on the boundary, 
$\th_1$ is no longer BRST invariant. Thus there are no boundary observables 
(except for the identity operator) associated with the D2 brane.

\subsection{The Boundary Operator Product Algebra}

First let us compute topological correlators for strings connecting
a brane with itself. Since in the D2-D2 case there is only the vacuum state,
the problem is non-trivial only in the D0-D0 case. 

\begin{figure}[b]
  \centerline{\epsfxsize=10cm\epsfbox{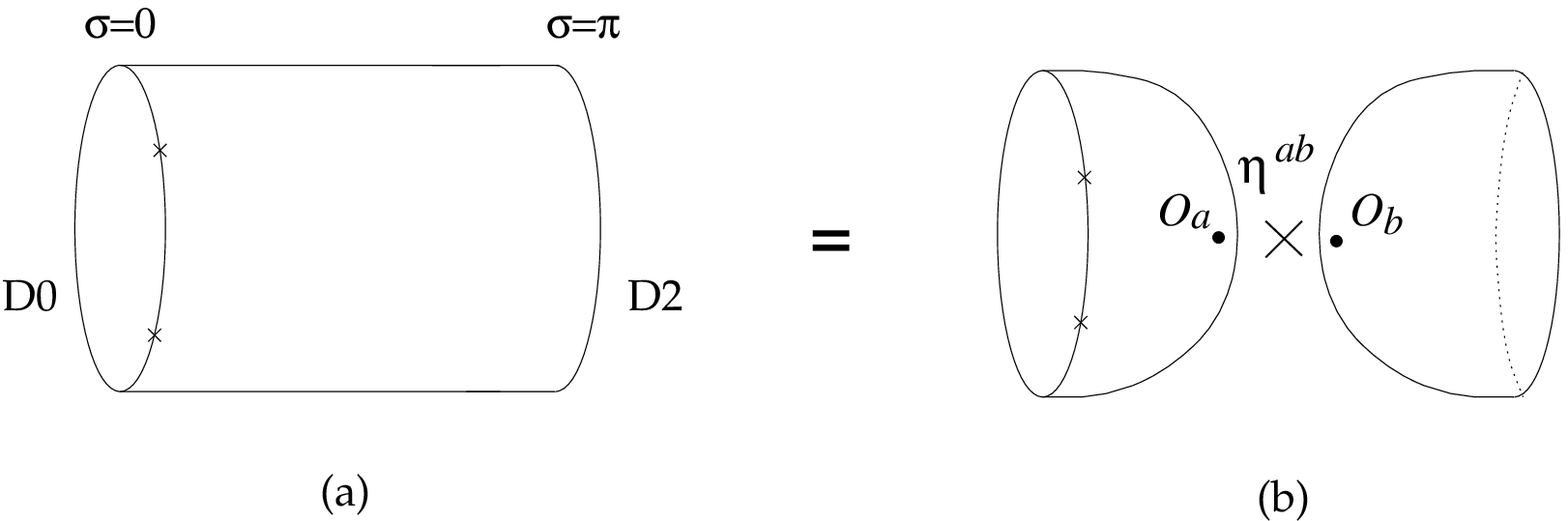}}
  \caption{}
  \label{fig:cyl}
\end{figure}

In order to compute disk correlators with products of $\th_1$ and $\th_2$ 
inserted on the circumference, we proceed as in Ref.~\cite{H}. 
We start with a world-sheet diagram which has the topology of a cylinder,
with the D0 and D2 boundary conditions imposed on the two boundary circles
(c.f. Fig.~\ref{fig:cyl}). On the D0 boundary there can be operator
insertions. Viewed in the open-string channel, this world-sheet diagram 
computes the one-loop amplitude
\begin{equation}
	\l\langle \B_1\B_1\cdots \B_r\r\rangle_{\rm cyl} = \Tr\l[(-)^Fe^{-\ep H}\B_1\B_2\cdots\B_r\r]
\end{equation}
where the $\B$'s are operators inserted at the D0 boundary. 
The trace on the RHS can be reduced to the Hilbert space of open-string zero 
modes by the standard argument. In the specific case at hand, the zero mode 
space is spanned by $|0\rangle$ and $\zeta^\dagger|0\rangle$ as described in 
Sec.~\ref{sec:d0d2}, and one easily obtains
\begin{eqnarray}
	\l\langle 1 \r\rangle_{\rm cyl} &=& 0,\no\\
	\big\langle \th_i \big\rangle_{\rm cyl} &=& 0,\no\\
	\big\langle \th_1\th_2 \big\rangle_{\rm cyl} &=& -\big\langle \th_2\th_1 \big\rangle_{\rm cyl} =- \frac{im}{2}.\no
\end{eqnarray}
More generally, one can compute
\begin{align}
	\big\langle (\th_1\th_2+\th_2\th_1)\cdot\B \big\rangle_{\rm cyl} = & -\frac{im}{2}\langle\B\rangle_{\rm cyl} & \forall \;\B,
	\label{eq:cyl_1}\\
	\big\langle \th_i^2 \cdot\B \big\rangle_{\rm cyl} = & 0 & \forall \;\B,\ \forall \;i.
	\label{eq:cyl_2}
\end{align}
{}From (\ref{eq:cyl_1}) and (\ref{eq:cyl_2}) one deduces the relations in the boundary 
operator product algebra for the D0-brane:
\begin{align}
	& \th_1\cdot\th_1 = \th_2\cdot\th_2 = 0,\\
	& \th_1\cdot\th_2 + \th_2\cdot\th_1 = -\frac{im}{2}.
	\label{eq:ope}
\end{align}
This is the Clifford algebra with two generators corresponding to the
quadratic form 
$$
-\frac{i}{2} \begin{pmatrix} 0 & m \\ m & 0\end{pmatrix}
$$ 
Up to a numerical factor, this matrix is the Hessian of $W$ in the basis $\theta_1,\theta_2$.
Thus one can state the result more invariantly by saying that the
boundary operator product algebra for the D0-brane is the Clifford algebra
$Cl(V,Q)$, where $V$ is target space of our LG model, and $Q$ is the
quadratic form given by the Hessian of $W$. 

Topological correlators on the disk can be inferred from the computation 
on the cylinder using factorization in the closed string channel~\cite{W:2d}. 
Namely, we insert a complete set of states in the closed string channel (cf. Fig.~\ref{fig:cyl}b) and rewrite
the cylinder amplitude as
$$\langle \B\cdot O_a\rangle_{\rm D0}\cdot\eta^{ab}\cdot\langle O_b\rangle_{\rm D2},$$
where $O_a$'s form a complete set of bulk operators, and $\eta$ is the (inverse) metric on the space of bulk 
operators defined via 
topological correlators on the sphere. The relative normalization is fixed by demanding the following relation for the D2-D2 
cylinder amplitude
$$\l\langle 1\r\rangle_{\rm D2-D2}=1.$$
In our case, the only bulk operator is the identity, therefore all disk correlators for the D0-brane simply coincide with 
the cylinder correlators.

Besides the algebra structure, another important datum is a non-degenerate inner product on the space
of endomorphisms. This inner product is determined by the two-point disk correlator and makes the endomorphism
algebra into a (non-commutative) graded Frobenius algebra. In our case the only non-vanishing inner products are
$$
\langle \theta_1\theta_2, 1\rangle=\langle 1,\theta_1\theta_2\rangle=\langle \theta_1,\theta_2\rangle=-\langle \theta_2,
\theta_1\rangle=-\frac{im}{2}.
$$
Note that the bilinear form corresponding to this product is even. In contrast, in the model $W=z^2$ the bilinear form
is odd and given by
$$
\langle 1,\theta\rangle=\langle \theta, 1\rangle = 1.
$$

So far we have determined the endomorphism algebra of the D0-brane (it is
isomorphic to $Cl(2,\CC)$) and the D2-brane (it is isomorphic to $\CC$).
Now we turn to the computation of compositions of morphisms between different
branes. 

We begin with the case when both branes are D2-branes. Let us denote
the D2-brane given by the equation $\Phi_1=0$ (resp. $\Phi_2=0$) by
$Y_2$ (resp. $Y_1$). It was shown in the previous section
that the vector space $Mor(Y_i,Y_j)$ is one-dimensional for all $i$ and
$j$. When $i=j$, this space is even, but for $i\neq j$ there is no 
canonical choice for the R-charge.  In other words, for $i\neq j$
the ``vacuum'' vector spanning $Mor(Y_i,Y_j)$ can equally well be 
regarded as even or odd. 
For reasons which will become clear later, we define $Mor(Y_1,Y_2)$
to be purely odd; since $Mor(Y_2,Y_1))$ is dual to $Mor(Y_1,Y_2)$, it
is also purely odd, while $Mor(Y_1,Y_2[1])$ is purely even. Here
$Y_2[1]$ denotes the shift of $Y_2$.

Let $\gamma_{12}$ and $\gamma_{21}$ be generators of $Mor(Y_1,Y_2)$ 
and $Mor(Y_2,Y_1)$, respectively.
Since the endomorphism algebra of a D2-brane is spanned by the identity
morphism, we only need to determine if $\gamma_{12}\cdot\gamma_{21}$ is zero 
or not. This product is evaluated by the disk amplitude with two
insertions of boundary-changing operators. By conformal invariance, this
is the same as the vacuum-vacuum transition amplitude for open strings
stretched between $Y_1$ and $Y_2$. Since there are no fermionic zero
modes in this case, this amplitude is non-zero. This means that
$\gamma_{12}\cdot\gamma_{21}= c\cdot id$ with $c\neq 0$.

This trivial computation implies that the even generator of 
$Mor(Y_1,Y_2[1])$ is an isomorphism. In physical language, $Y_1$ is isomorphic to
the anti-brane of $Y_2$.

There remain compositions of morphisms involving both D2-branes and
the D0-brane. In view of the previous paragraph, it is sufficient
to consider morphisms between $Y_1$ and the D0-brane. 
No new computations are actually required,
the result being fixed by general properties of topological string theory~\cite{MS,Laz}. First of
all, we note that by cyclic symmetry of topological correlators computing
compositions of morphisms from D2 to D0 and back (or the other way around)
is equivalent to computing how the endomorphism algebra of the D0-brane
acts on the space of morphisms from D2 to D0. In more detail, we have a non-degenerate pairing
$$
Mor(Y_1,D0)\times Mor(D0,Y_1)\ra \CC
$$
given by the path-integral on an infinite strip. (This paring is odd in our case, because there is a single
fermionic zero mode.) Similarly, we have an even non-degenerate pairing
$$
Mor(D0,D0)\times Mor(D0,D0)\ra\CC.
$$ 
Thus computing the product map
$$
Mor(D0,Y_1)\times Mor(Y_1,D0)\ra Mor(D0,D0)
$$
is the same as computing the map
$$
Mor(Y_1,D0)\times Mor(D0,D0) \ra Mor(Y_1,D0).
$$
Furthermore, in our case $Mor(D0,D0)$ is isomorphic to $Cl(2,\CC)$, and
we know from the previous
section that $Mor(Y_1,D0)$ is two-dimensional. 
The $\ZZ_2$
graded algebra $Cl(2,\CC)$ has a unique representation on $\CC^2$, up to a 
flip of parity (up to isomorphism, it is given by any two Pauli matrices).
Since in string theory the parity of morphisms is not canonically fixed
anyway, we conclude that the module structure of $Mor(Y_1,D0)$
is completely determined, up to the unavoidable ambiguity in the overall
parity. This in turn determines the composition of morphisms going from
D0 to D2 and back.

\section{B-branes in general massive LG models}
\label{sec:D}

\subsection{Generalities}\label{secDgen}
We now turn to massive Landau-Ginzburg models which involve more than two 
fields. Without loss of generality, we may assume that the superpotential on $\CC^n$ is given by
\begin{equation}\label{generalW}
W=z_1^2+\ldots +z_n^2.
\end{equation}
We can construct examples of B-branes in this LG model for any $n$, using 
the results of the previous section. For $n=2k$, $k\in\ZZ$, we consider an 
equivalent superpotential
$$
W=z_1 z_2 +z_3 z_4+\ldots + z_{2k-1} z_{2k}.
$$
Since it is a sum of $k$ copies of the superpotential $W=xy$, we can construct
a B-type boundary condition by picking $k$ arbitrary B-type
boundary conditions for the latter model, and tensoring them. For example,
if we take all boundary conditions to be D0-branes, the tensor product state
will also be a D0-brane, and its endomorphism algebra will be $Cl(2k,\CC)$. If
we take all boundary conditions to be D2-branes, the tensor product boundary
state will be a D($2k$)-brane, and its endomorphism algebra will be $\CC$.

Similarly, for $n=2k+1$ we consider an equivalent superpotential
$$
W=z_1 z_2 +z_3 z_4+\ldots + z_{2k-1} z_{2k}+z_{2k+1}^2.
$$
Clearly, B-branes for this LG model can be constructed by taking tensor
product of $k$ boundary states for the LG model with $W=xy$ and a boundary
state for the LG model with $W=z^2$. In this way one obtains B-branes of dimension
up to $2k$. It is easy to see that the endomorphism algebra of the D0-brane will be
isomorphic to $Cl(2k+1,\CC)$, while the endomorphism algebra of the
D($2k$)-brane will be isomorphic to $Cl(1,\CC)$.

More generally, one can explicitly construct all B-branes which correspond
to linear subspaces of the critical level set $W=0$. Since $W$ is quadratic, these are the
same as linear subspaces isotropic with respect to the bilinear form $Q$. Classification of
such isotropic subspaces is well known~\cite{GH}. The maximal
dimension of an isotropic subspace is $[n/2].$ For $n$ odd, there is a single
irreducible family of isotropic subspaces of maximal dimension parametrized by
$(n-1)(n+1)/8$ parameters. For $n$ even, there are two irreducible families 
of isotropic subspaces of maximal dimension parametrized by $n(n-2)/8$
parameters. Any isotropic subspace lies in one of the maximal isotropic
subspaces. It is straightforward to compute morphisms and their 
compositions (i.e the spectrum and topological correlators) between all linear
B-branes. In the next subsection we discuss in some detail
the results for the case $n=3$, when the LG superpotential has the
form $W=xy+z^2$. Then we will describe the general case.

Note that in principle there could also be B-branes corresponding to
non-linear boundary conditions (e.g. non-linear submanifolds of the
quadric $W=0$). Such B-branes are hard to study directly. In what follows 
we shall focus on linear boundary conditions. 

\subsection{The LG model with the superpotential $W=xy+z^2$}
\label{sec:W3}

Maximal isotropic linear subspaces on the quadric surface $W=0$ are complex lines, 
and there is a single irreducible family of them.
This family is parametrized by $\CC\PP^1$ as follows:
$$
\mu x+\nu z=0,\quad \mu z -\nu y = 0,
$$
where $[\mu:\nu]$ are homogeneous coordinates on $\CC\PP^1$. Any two
distinct lines in the family intersect at a single point ($x=y=z=0$).

Using a linear change of basis in the target space which preserves
$W$, one can always map any line in the above family to the line
$x=z=0$. For the brane $x=z=0$ we already know that the endomorphism algebra
is isomorphic to $Cl(1,\CC)$, and since linear changes of variables
preserving the superpotential are invariances of the topological LG
model, we conclude that the same is true for any D2-brane in the above family.

Next we consider morphisms between different lines in the family.
Clearly, there are no bosonic zero modes, so the space of
morphisms will be spanned by the ``vacuum'' state and its fermionic
excitations with zero energy. As remarked in the previous section,
only some components of $\theta$ have a chance to be BRST-non-trivial
boundary observables. Thus all BRST-invariant states can be obtained by
acting by some components of $\theta$ on the vacuum state.

Let $V\simeq\CC^3$ be the target space of our LG model, and let $U_1$
and $U_2$ be two distinct lines in $V$ isotropic with respect to the quadratic form $W$.
The corresponding B-branes will be denoted $Y_1$ and $Y_2$.
Let us look at the $\theta$-field restricted to the boundary of the world-sheet
which is mapped to $U_1$. We can regard $\theta_i$ as basis elements of $V$. BRST transformations
are
$$
\delta \theta_i = Q_{ij}\phi^j.
$$
On the boundary the vector with components $\phi^j$ can be an arbitrary element of $U_1$.
It follows that BRST-invariant components of $\theta$ must be orthogonal to $U_1$ with respect to
the form $Q$.
We denote the orthogonal subspace by $U_1^{\perp}$. Of course, since $U_1$ is isotropic,
we have an inclusion $U_1\subset U_1^\perp$. Similarly, 
BRST-invariant fermionic
fields on the $U_2$-boundary are parametrized by elements of $U_2^{\perp}\supset U_2$.
The total space of BRST-invariant fermionic fields is
$$
U_1^{\perp}\op U_2^{\perp}= V.
$$
However, not all of these are non-zero. Neumann boundary conditions
plus supersymmetry imply that the components of $\theta$ along $U_1\op U_2$ vanish.
Thus non-trivial BRST invariant fermionic zero modes are parametrized
by elements of the quotient space
$$
V/(U_1\op U_2).
$$
This space is one-dimensional.
Thus there is a single fermionic
zero mode, and the space of morphisms between two different lines is isomorphic to its exterior
algebra (as a $\ZZ_2$-graded vector space). That is, $Mor(Y_1,Y_2)$ has one-dimensional even subspace, and one-dimensional
odd subspace.

Composition of morphisms between two distinct
lines is fixed by consistency considerations. If $Y_1$ and $Y_2$ are
any two lines, then $Mor(Y_1,Y_2)$ must be a left module over
$Mor(Y_1,Y_1)\simeq Cl(1,\CC)$ and right module over $Mor(Y_2,Y_2)\simeq
Cl(1,\CC)$. There is only one such module
of dimension two: the Clifford algebra itself, regarded as 
a bi-module over itself. Together with various parings given by the 2-point correlators, 
this fixes the structure of correlators involving any
two D2-branes. In particular, it is easy to see that the element in $Mor(Y_1,Y_2)$
corresponding to the identity element in $Cl(1,\CC)$ is invertible. 
This means that any two lines give isomorphic objects in the category
of B-branes. 

Similar arguments can be used to determine boundary correlators
involving both D2 and D0. As explained above, the endomorphism algebra
of the D0-brane is isomorphic to $Cl(V,Q)\simeq Cl(3,\CC)$. As for the space of morphisms
between a D2-brane and D0-brane, it is 4-dimensional, with two-dimensional
even subspace and two-dimensional odd subspace. Indeed, since all D2-branes
are isomorphic, it is sufficient to consider morphisms from 
the D2-brane $x=z=0$. This D2-brane is the tensor
product of the D2-brane in the LG model $W=xy$ and the D0-brane in the
LG model $W=z^2$. Hence the computation of the space of morphisms and their compositions
is reduced to the one we have performed in the previous section.

\subsection{The LG model with the superpotential $W=z_1^2+\cdots +z_n^2$}

The above arguments can be easily generalized to arbitrary $n$.
We shall consider only linear boundary conditions.
Let the target space be $V\simeq\CC^n$, let $W$ be a non-degenerate
quadratic function on $V$, and let $Q\in Sym^2(V^*)$ be its Hessian.
A B-brane is a linear subspace $U\subset V$ which is isotropic with respect
to $Q$. As mentioned above, $k=\dim_\CC U$ is less or equal to $[n/2]$.

Using linear changes of variables, we can bring $W$ to the standard
form, and $U$ to the subspace given by $z_1=\ldots=z_{n-k}=0$. Such a
D(2k)-brane is a tensor product of $k$ copies of D2-branes in the model
$W=xy$, $[n/2]-k$ copies of the D0-brane in the model $W=xy$,
and, for $n$ odd, one copy of a D0-brane in the model $W=z^2$.
It follows that the space of endomorphisms has dimension 
\begin{equation}\label{dimend}
dim_\CC \End(D(2k))=2^{n-2k},
\end{equation}
and is isomorphic as a $\ZZ_2$-graded algebra to the Clifford algebra with 
$n-2k$ generators. In particular, the algebra of endomorphisms
of a D-brane of maximal possible dimension is isomorphic to $\CC$
or $Cl(1,\CC)$ depending on whether $n$ is even or odd, while the
the endomorphism algebra of the D0-brane is isomorphic to $Cl(V,Q)\simeq Cl(n,\CC)$.

Next let us discuss morphisms between two different B-branes. Let
$U_1$ and $U_2$ be isotropic linear subspaces corresponding to B-branes $Y_1$ and $Y_2$.
The same arguments as in the previous subsection tell us that the
space of fermionic zero modes can be identified with
$$
(U_1^\perp\op U_2^\perp)/(U_1\op U_2).
$$
The space of morphisms is isomorphic as a graded vector space
to the exterior algebra of this vector space (up to an overall flip of parity).
It is easy to see that the dimension of the space of zero modes is given by $n-k_1-k_2$,
where $k_i=dim\ U_i$. Therefore the dimension of the space of morphisms is given by
$$
2^{n-k_1-k_2}.
$$
In particular, in the case $U_1=U_2$ we recover the result Eq.~(\ref{dimend}) obtained by other
means.

Let us give a few examples. First, let $n$ be even, and $U_1$ and $U_2$
be distinct maximal isotropic subspaces. Then $U_i^\perp=U_i$ for $i=1,2$,
and there are no zero modes. This means that the space of morphisms
between any two maximal isotropic subspaces is one-dimensional. As usual,
the R-charge assignment is ambiguous, but it is natural to require the
R-charge to vary continuously as one varies $U_i$. Since in the case
$U_1=U_2$ the space of endomorphisms is even and isomorphic 
to $\CC$, this implies that for any two maximal isotropic subspaces
in the same irreducible family the space of morphisms is even and isomorphic to $\CC$.
We will fix the remaining ambiguity by saying that the space of
morphisms between two maximal isotropic subspaces in {\it different} irreducible
families is odd. The reason for such a convention will be explained in the next section.

The fact that for even $n$ there are no fermionic zero modes for open strings 
connecting two maximal isotropic subspaces implies that the vacuum-vacuum
transition amplitude is non-zero in this sector. This is equivalent to
saying that the composition of non-zero morphisms between two maximal 
B-branes is a non-zero multiple of the identity endomorphism. 
If these two B-branes are in the same irreducible family, this means
that they represent isomorphic objects in the category; if they are in 
different irreducible families, then the interpretation is that they
are isomorphic up to a shift.

If $n$ is odd, and $U_1$ and $U_2$ are maximal linear subspaces,
then there is a single fermionic zero
mode. Thus $Mor(Y_1,Y_2)$ is two-dimensional, with one-dimensional even
and one-dimensional odd subspaces. The composition of morphisms going between
two maximal B-branes is fixed by consistency requirements. Namely,
the space of morphisms must be a graded bi-module over $Cl(1,\CC)$ (the
endomorphism algebra of a single B-brane), and there is only one such
graded bi-module of dimension two: $Cl(1,\CC)$ itself. Furthermore, there is
an odd non-degenerate pairing 
$$
Mor(Y_1,Y_2)\times Mor(Y_2,Y_1)\ra \CC,
$$
which is invariant with respect to both actions of $Cl(1,\CC)$ in an obvious
sense. Up to isomorphism, there is only one such pairing, namely
$$
\langle a,b\rangle=tr(ab),
$$
where $tr$ is defined by
$$
tr(1)=0,\quad tr(\theta)=1.
$$
Together with the module structure of $Mor(Y_1,Y_2)$, this pairing
determines the composition of morphisms going between any two lines.
As in the case $W=xy+z^2$, it is easy to see that the morphism corresponding
to the identity element of $Cl(1,\CC)$ is invertible, and therefore any two maximal B-branes are isomorphic. 

Our third example is the case $U_1=U_2=0$, that is, the case of the D0-brane.
The space of zero modes coincides with $V$, and the space of endomorphism
is isomorphic to $\wedge^*V$ as a $\ZZ_2$-graded vector space ($V$ is regarded as odd). This agrees
with an independent argument of subsection~\ref{secDgen}. There we also showed that
the algebra of endomorphisms is isomorphic to $Cl(V,Q)$.

Our fourth and final example is the case when $U_1$ is an arbitrary 
isotropic subspace
of dimension $k\leq [n/2]$,
and $U_2=0$. In other words, the second brane is the D0-brane. Then
the space of zero modes is $V/U_1$. Its dimension is $n-k$, and
therefore the space of open strings stretched between a maximal linear
subspace and the D0-brane has dimension $2^{n-k}.$ The space of morphisms has the
structure of a graded module over the endomorphism algebra of the D0-brane, which is
isomorphic to $Cl(V,Q)$. If we neglect the grading, there is a unique such module,
which is a sum of $2^{[n/2]-k}$ irreducible (spinor) modules.

\section{B-branes and Twisted Complexes}\label{sec:E}

\subsection{Kontsevich's proposal}\footnote{The content of this subsection was explained
to us by Maxim Kontsevich.}

Let us recall how to define the derived
category of coherent sheaves on a smooth affine variety $X$ following Ref.~\cite{BK}.
Let $Coh(X)$
be the category of coherent sheaves on $X$, or equivalently the category
of finite modules over the coordinate ring $\O_X$ of $X$. We define $C(X)$ 
to be a category whose objects are bounded $\ZZ$-graded complexes of projective 
objects of $Coh(X)$.
Equivalently, we can think about the coordinate ring of $X$ as a 
differential graded algebra (dg-algebra) which is concentrated in degree zero and has
a trivial differential; then objects of $C(X)$ are differential graded
modules (dg-modules) over this dg-algebra such that all homogeneous components are 
projective $\O_X$-modules, and all but a finite number of homogeneous
components are trivial.
Morphisms in $C(X)$ are morphisms of these dg-modules regarded
simply as $\O_X$-modules (i.e. morphisms do not necessarily 
preserve the grading or respect the differentials). 
Groups of morphisms in the category $C(X)$ are 
naturally $\ZZ$-graded and have a natural differential of degree $1$. 
For example, closed morphisms of degree 0 in the category $C(X)$ are 
ordinary morphisms of complexes (the ones which preserve the grading and 
commute with the differentials), while exact morphisms of degree 0 are
morphisms of complexes which are homotopic to zero.
Thus $C(X)$ is a dg-category. 

There is a general way to make a triangulated 
category out of 
any dg-category~\cite{BK}. One takes the category of ``twisted objects'' 
of the dg-category, which is again a dg-category, and then passes to degree-0 
homology, i.e. replaces groups of morphisms with their degree-0 homology.
In the present case, since we are working with complexes of projective modules,
it is not necessary to consider twisted objects, and
one can simply apply the functor $H^0$ to $C(X)$. 
The resulting triangulated category is simply the homotopy category
of bounded complexes of projective $\O_X$ modules, and it is well known that
it is equivalent to the bounded derived category of $Coh(X)$ 
(see e.g. Ref.~\cite{GM}). Alternatively, one can apply to $C(X)$ the functor $H^*$.
This gives a graded category which is the completion of $D^b(Coh(X))$ with
respect to the shift functor. As discussed in Section~\ref{sec:A}, the latter alternative
conforms better to physical conventions.

Now we can formulate Kontsevich's proposal rather simply. Let $X$ be
a smooth affine variety, and $W$ be a holomorphic function on $X$ (the superpotential),
whose critical set is compact. Let $W_0\in\CC$ be a critical value of
$W$. First, since
in the presence of the superpotential morphisms between B-branes
are $\ZZ_2$-graded, we will have to use $\ZZ_2$-graded complexes in order
to construct the analogue of $C(X)$.
Second, we deform our $\ZZ_2$-graded complexes of projective
modules by asking that the composition of two successive morphisms be
equal to $W-W_0$, instead of zero. Thus objects of the
deformed category $C(X,W,W_0)$ are pairs of finitely generated projective $\O_X$-modules
$E_0,E_1$ and morphisms $d_0:E_0\ra E_1$ and $d_1:E_1\ra E_0$ such
that
$$
d_1 d_0= W-W_0,\quad d_0 d_1= W-W_0.
$$
We can regard the pair $(E_0,E_1)$ as a $\ZZ_2$-graded $\O_X$-module, 
and $(d_0,d_1)$
as an odd endomorphism $d_E$ of this module whose square is $W-W_0$
(``twisted differential'').
Morphisms in this category are defined as (ungraded) morphisms of 
the corresponding $\O_X$-modules.
They have a natural $\ZZ_2$-grading, and a natural differential. 
The differential on $Mor(E,F)$ is defined as
$$
D\phi=\phi d_E +(-1)^{\rm deg}\d_F \phi.
$$
Here $(-1)^{\rm deg}:Mor(E,F)\ra Mor(E,F)$ acts as $1$ on the even component and as $-1$
on the odd component.
It is easy to see that $D:Mor(E,F)\ra Mor(E,F)$ is an odd operator 
whose square is zero. Thus $C(X,W,W_0)$ is
a differential $\ZZ_2$-graded category. In what follows the term ``dg-category'' (resp. ``graded category'')
will refer to a differential $\ZZ_2$-graded category (resp. $\ZZ_2$-graded category), unless specified otherwise.

Applying to $C(X,W,W_0)$ the functor $H^*$, we obtain
a graded category, which is proposed to be equivalent to the 
category of B-branes corresponding to the critical value $W_0$.
One can show that all spaces of morphisms in this category are finite-dimensional,
provided the critical set of $W$ is compact.

An unsatisfactory feature of this construction is that one needs to
use complexes of projective modules, instead of general complexes. 
This causes problems if one
tries to extend the definition from affine varieties to algebraic ones. There is a way
to repair this defect~\cite{Kpriv}, but we will not try to explain this
more complicated definition in this paper.

\subsection{A physical derivation of Kontsevich's proposal}

In this subsection we give a physical argument supporting the identification of B-branes with objects
of the category $C(X,W,W_0)$. 
Our argument is modelled on those in Refs.~\cite{AL,HMcG,Hetal}, where it was explained why complexes of locally
free sheaves on a Calabi-Yau manifold can be thought of as B-branes. For our purposes, it is
sufficient to consider $\ZZ_2$-graded complexes. Then the argument of Refs.~\cite{AL,HMcG,Hetal} can be
summarized as follows. Consider a pair of locally free sheaves (i.e. holomorphic vector bundles) $E_1$ and $E_2$.
We already know that $E_1$ and $E_2$ can be thought of as B-branes, i.e. as topological boundary conditions
for a topological sigma-model. The same goes for 
$E_1\op E_2$ and $E_1\op E_2[1]$. (In the physical setting, one has additional data, such as Hermitian metrics on $E_1$ and
$E_2$ and compatible connections.)  
Now we can deform the boundary
condition corresponding to $E_1\op E_2[1]$ by adding a boundary term to the action which depends on a pair of sections
$F\in Hom(E_1, E_2)$ and $G\in Hom(E_2,E_1)$ (the ``tachyons''). 
In order to preserve BRST invariance, one has to require that $F$ and
$G$ be holomorphic, and $FG=0$, $GF=0$. This deformed boundary condition corresponds to a $\ZZ_2$-graded complex
$$
\begin{CD}
E_1 @>F>> E_2 @>G>> E_1
\end{CD}
$$
One expects (although there is no iron-clad argument) that any B-brane is isomorphic to a B-brane of this kind.
This provides a physical explanation for the relation between complexes of locally free sheaves on a Calabi-Yau
and B-branes.

In the case of a LG model, locally free coherent sheaves on $X$ are not valid B-branes, because their support
is the whole $X$, and $W$ is not constant on $X$. Technically, the problem occurs because the BRST variation of the
bulk action contains a non-vanishing $W$-dependent term which is a total derivative on the world-sheet. This is the so-called
Warner problem~\cite{Warner}.
The sum $E_1\op E_2[1]$ is not a B-brane either. However, we can
try to add a boundary term to the action of the sigma-model so that BRST invariance is restored. We take the
same term as in the case $W=0$. As in Refs.~\cite{AL,HMcG,Hetal}, we have two holomorphic sections $F$ and $G$ to play with.
As we show below, the condition of BRST-invariance is modified to $FG=W-W_0$, $GF=W-W_0$, where
$W_0$ is a constant. This shows that any object of $C(X,W,W_0)$ corresponds to a B-brane.

Now let us work out the BRST-invariance conditions and show that $F$ and $G$ must satisfy the constraints stated above.
For simplicity, let us assume at first that both $E_1$ and $E_2$ are line bundles. The boundary Lagrangian is taken to be
\begin{equation}
        \L_b = \frac{i}2\l(\bar\ga D_\tau\ga +\psi^i\partial_iF\ga+
        \psi^{\bar{i}}\partial_{\bar{i}}\bar{G}\ga\r) -
        \frac14\l(\bar{F}F+\bar{G}G\r) \;+\; {\rm h.c.}\no
\end{equation}
Here $\gamma$ is a complex fermion living on the boundary, $\psi^i = \psi_+^i+\psi_-^i$ is the restriction of a bulk
fermionic field to the boundary, and $F=F(\phi),G=G(\phi)$ are holomorphic sections of $Hom(E_1,E_2)$ and $Hom(E_2,E_1)$, 
respectively. They depend on the fields $\phi^i$ restricted to the boundary. The fermion $\gamma$ takes values in $Hom(E_2,E_1)$,
and the covariant derivative $D_\tau$ along the boundary makes use of the unitary connections on $E_1$ and $E_2$.
The boundary Lagrangian is manifestly gauge-invariant. If we set $F=G=0$, we get the usual path-integral representation of
the parallel transport operator in the bundle $E_1\op E_2^*$~\cite{H}. For non-zero $F$ or $G$ we get a deformation of the usual
boundary condition. In the special case $G=0$ we get the boundary Lagrangian used in Ref.~\cite{H}.

We postulate the following supersymmetry transformations for $\gamma$: 
$$\delta\ga\;=\; i\ep\bar{F}-i\bep{G}.$$
Here $\ep$ and $\bep$ are regarded as independent complex Grassmann variables.
We also note that $F$ and $G$ transform as follows:
\begin{eqnarray}
        \delta{F} &=& \ep\psi^i\partial_iF\no\\
        \delta{G} &=& \ep\psi^i\partial_iG\no
\end{eqnarray}
BRST transformations are obtained by setting $\ep=0$. One can check that
the BRST variation of the boundary Lagrangian is given, up to a total derivative, by
\begin{equation}
\delta\L_b \;=\; -\frac12\bep\psi^i\partial_i(FG).\no
\end{equation}
On the other hand, the BRST variation of the bulk action is a boundary term given by
\begin{eqnarray}
        \delta S_0 &=& \int\d\tau\;\bar{\ep}\l[g_{i\bj}\partial_0\phi^i
        \l(\psi_-^\bj-\psi_+^\bj\r)+g_{i\bj}\partial_1\phi^i
        \l(\psi_-^\bj+\psi_+^\bj\r)\r.\no\\
        && \l. +\,\frac{i}{2}\l(\psi_-^i+\psi_+^i\r)
        \partial_i{W}\r].\no
\end{eqnarray}
The first two terms in the bulk variation are standard and vanish when the standard Neumann boundary conditions are imposed
on $\phi^i$ and $\psi^i_\pm$. The last term is the Warner term~\cite{Warner}. Obviously,
in order for the variation of the boundary Lagrangian to cancel the Warner term, we need to require
$$F(\phi)G(\phi) \;=\; {i}(W(\phi)+const).$$
One can get rid of the factor $i$ by redefining $G\ra iG$. 
This implies that instead of an ordinary complex of holomorphic vector bundles we are dealing with an object of
$C(X,W,W_0).$ 

It is straightforward to generalize the construction to the higher-rank case. The fermion $\gamma$ still takes values in
$Hom(E_2,E_1)$, which means that it is a matrix of size ${\rm rank}(E_1)\times {\rm rank}(E_2)$. 
In order for the path-integral over $\gamma(\tau)$ to 
reproduce a path-ordered exponential in the representation of the gauge group of dimension 
${\rm rank}(E_1)+ {\rm rank}(E_2)$, one
needs to insert a projector onto the sector where the total fermion number (including the boundary contribution from $\gamma$)
is equal to $0$ or $1$~\cite{HMcG,Hetal,GJ2}. The rest of the argument is unchanged. The conditions of BRST-invariance
now read
$$
FG=i(W+const),\quad GF=i(W+const).
$$
The two conditions arise by requiring that the BRST variation of the bulk term be cancelled on both boundaries of the world-sheet.
By taking the trace of these two equations and comparing them, one infers that the ranks of $E_1$
and $E_2$ are in fact the same, and the constant terms in the equations are also the same. It follows that any object of
$C(X,W,W_0)$ is a B-brane.

One can also check that the total BRST charge is nilpotent. Indeed, it is easy to see that the square of the bulk contribution
to the BRST charge is equal to
$$
Q_0^2 \;=\; \frac12\{Q_0, Q_0\} \;=\; -i W|_{\partial\Sigma}.
$$
On the other hand, the boundary supercharge coming from one of the two boundaries is given by
$$Q_b \;=\; -iF\ga+iG\bar{\ga}.$$
Canonical quantization yields
$$\{\ga,\bar{\ga}\} \;=\; 1$$
and therefore 
\begin{equation}
        Q_b^2 \;=FG\no
\end{equation}
It is also easy to check that $Q_b$ and $Q_0$ anti-commute (the holomorphicity of $F$ and $G$ is important here.)
Hence the sum of $Q_0$ and the two boundary supercharges is nilpotent.

\subsection{Checking Kontsevich's proposal}\label{sec:konts}

We start with the case $X=\CC$, $W=z^2$, where there is only a 
D0-brane to worry about. In the absence of the superpotential, D0-brane on $\CC$ is associated
with the structure sheaf of a point, which has a two-term projective
resolution
$$
\begin{CD}
\O @>z>> \O.
\end{CD}
$$
If we pass to $\ZZ_2$-graded complexes, we obtain the following
dg-module:
$$
\begin{CD}
\O\op\O @>z\sigma_->> \O\op\O,
\end{CD}
$$
where 
$$
\sigma_-=\begin{pmatrix} 0 & 0\\ 1 & 0\end{pmatrix},
$$
and it is understood that the module $\O\op\O$ has the obvious grading.
If we turn on the superpotential, we need to deform the differential so that
its square be equal to $W=z^2$. It is clear how to do this: simply
consider the object
\begin{equation}\label{D0object}
\begin{CD}
\O\op\O @>z\sigma_1>> \O\op\O,
\end{CD}
\end{equation}
where
$$
\sigma_1=\begin{pmatrix} 0 & 1\\ 1 & 0\end{pmatrix}.
$$
This is our candidate object for the D0-brane at $z=0$. As a check, let us
compute its endomorphism algebra, following Kontsevich's prescription.
In the category $C(X,W,W_0)$, the algebra of endomorphisms is 
$$
\End_\O(\O)\ot_\CC Mat(2,\CC)\simeq Mat(2,\CC[z]).
$$
The $\ZZ_2$-grading is the natural grading on $2\times 2$ matrices
(diagonal elements are even, off-diagonal elements are odd). 
The differential acts on this graded vector space as follows:
$$
D: \begin{pmatrix} A & B\\C & D\end{pmatrix}\mapsto \begin{pmatrix}
(B+C)z & (A-D)z \\ -(A-D)z & (B+C)z\end{pmatrix}.
$$
Here $A,B,C,D$ are elements of $\End_\O(\O),$ i.e. 
simply polynomials in $z$.
Computing $H^*$, we find that this abelian group is isomorphic to the
group  of complex matrices of the form
$$
\begin{pmatrix} \xi & \eta \\ -\eta & \xi\end{pmatrix}.
$$
Multiplicative structure is given by matrix multiplication. Clearly, this
algebra is generated over $\CC$ by the identity and an odd matrix 
$$
\sigma_2=\begin{pmatrix} 0 & -i \\ i & 0\end{pmatrix},
$$
which squares to identity.
This agrees with the endomorphism algebra of the D0-brane in the LG model $W=z^2$~\cite{H}.

Now let us discuss B-branes in the LG model $W=xy$. Using the same
reasoning as above, it is easy to guess that the D2-brane given by the
equation $x=0$ should correspond to the object
$$
\begin{CD}
\O\op\O @>>> \O\op\O,
\end{CD}
$$
where the map is defined as
$$
\begin{pmatrix} 0 & y\\ x & 0\end{pmatrix}.
$$
Similarly, the D2-brane given by the equation
$y=0$ should correspond to 
$$
\begin{CD}
\O\op\O @>>> \O\op\O
\end{CD}
$$
with the twisted differential
$$
\begin{pmatrix} 0 & x\\ y & 0\end{pmatrix}.
$$
The group of endomorphisms of the former object in the category
$C(X,W,W_0)$ is
$$
\End_\O(\O)\ot Mat(2,\CC)\simeq Mat(2,\CC[x,y]),
$$
with the differential which acts as follows:
$$
\begin{pmatrix} A & B\\C & D\end{pmatrix}\mapsto \begin{pmatrix}
Bx+Cy & (A-D)y \\ -(A-D)x & Bx+Cy\end{pmatrix}.
$$
The homology is readily computed; the result is that it is 
spanned by the identity matrix.
Thus the algebra of endomorphisms in the derived category is isomorphic to 
$\CC$. This agrees with the computation in Section~\ref{sec:Cp}. 
Of course, for
the other D2-brane we get the same result. Finally, in order to compute
morphisms between the two D2-branes, we note that one is a shift of the
other.\footnote{In physical terms, this means that the D2-brane with the
equation $x=0$ is isomorphic to the anti-brane for the D2-brane with the 
equation $y=0$.} 
Thus the space of morphisms is the space of endomorphisms with
gradings reversed, i.e. it is spanned by the identity matrix regarded as odd.
Composing two such odd morphisms going in the opposite directions we get 
the identity endomorphism. This agrees with the computations in 
Section~\ref{sec:Cp} and explains why we declared the space of morphisms between
two different D2-branes to be odd.

Now let us discuss the D0-brane. Consider the
direct sum of objects corresponding to D2-branes with equations $x=0$ and
$y=0$. It is easy to see that its algebra of endomorphism is the
Clifford algebra $Cl(2,\CC)$, so we propose that this object corresponds
to the D0-brane. It is easy to check that morphisms to and from other objects
agree with our computations in Section~\ref{sec:Cp}.

Finally, we propose an object of $H^*(C(X,W,W_0))$ corresponding to the 
D0-brane in the free massive LG model with $n$ fields. If we bring the superpotential
to the standard form Eq.~(\ref{generalW}), then we can simply tensor $n$ copies of the object
Eq.~(\ref{D0object}). Consequently, the endomorphism algebra will also be the graded tensor product
of $n$ copies of $Cl(1,\CC)$, which is isomorphic to $Cl(n,\CC)$. More invariantly, let
$V=X$ be the complex vector space
whose coordinates we denoted by $z_i$, let $e_i, i=1,\ldots,n,$ be the 
corresponding basis in $V$, let $e_i^*, i=1,\ldots, n$ be the dual basis
in $V^*$, and let $Q\in Sym^2(V^*)$ be
the Hessian of $W$. We start with the $\ZZ_2$-graded
version of the Koszul resolution of the point at the
origin:
$$
\Omega^{even}\ra \Omega^{odd},
$$
where $\Omega^i=\wedge^i V\ot \O_V$, and the differential is induced by
the wedge product with $z_i e_i$ (we use Einstein's convention of summing
over repeating indices). Now we modify the differential so
that its square be $W$ instead of zero. The obvious guess is
$$
d=z_i\left(e_i+\frac{1}{2}Q_{ij}i(e_j^*)\right),
$$
where $i(u), u\in V^*$, denotes the interior product with an element of $V^*$.
Using the identity
$$
\left\{e_i,i(e_j^*)\right\}=\delta_{ij},
$$
one can easily check that $d^2=W$. Note that $d$ is essentially the Fourier transform of the
Dirac operator, if we identify $\Omega^{even}$ and $\Omega^{odd}$ with
spinor bundles.

\subsection{B-branes, Clifford modules, and Koszul duality}
\footnote{The content of this subsection was
explained to us by Alexander Polishchuk.}
We regard the above computations as a convincing check of Kontsevich's
proposal for massive LG models. In this subsection, we
would like to address the following three questions. First, how do we 
match B-branes with objects in the
category $H^*(C(X,W,W_0))$ if $n>2$? (As explained in Section~\ref{sec:B}, it
is sufficient to consider the case $X=\CC^n$, $W$ quadratic non-degenerate,
and $W_0=0$.) Second, is
there an efficient method to compute morphisms in the category
$H^*(C(X,W,W_0))$? Third, assuming the validity of Kontsevich's proposal,
what do we learn about B-branes in massive LG models? That is, is there
a simpler way to describe $H^*(C(X,W,W_0))$?

To answer these questions, we will define a functor from the category
of B-branes to the category of finite-dimensional $\ZZ_2$-graded modules 
over the Clifford algebra $Cl(n,\CC)$. Let us denote this category
$\ClMod(n)$. (As usual, we allow both even and odd morphisms; thus $\ClMod(n)$ is
a $\ZZ_2$-graded category.) Since we set $X=\CC^n$, $W_0=0$, and $W$ is quadratic and non-degenerate, 
the category $H^*(C(X,W,W_0))$ really depends only on $n$; we will denote this category
$\K(n)$ for short. There is a further functor from $\ClMod(n)$ to
$\K(n)$. Composing these two functors gives a way to associate objects of $\K(n)$
to B-branes. In fact, as explained below, the second functor is an equivalence of categories
which implies that we 
can calculate morphisms in $\ClMod(n)$ instead of $\K(n)$. This equivalence is a cousin
of the much-studied Koszul duality for quadratic algebras (see below).
The structure of $\ClMod(n)$ is quite simple: any object is a direct sum of
irreducible objects (spinor modules), and there is one or two non-isomorphic
irreducible objects, depending on whether $n$ is odd or even. 
Thus we have a completely explicit description of $\ClMod(n)$, and 
therefore, by Koszul duality, of $\K(n)$. Assuming the validity
of Kontsevich's conjecture, this amounts to a solution of topological open
string theory for any massive LG model. 

One can associate a $\ZZ_2$-graded Clifford module to a B-brane as follows. 
For any B-brane $Y$, consider the graded vector space $M(Y)=Mor(D0,Y)$. Since
$Mor(D0,D0)$ is isomorphic to $Cl(n,\CC)$ as a graded algebra, $M(\cdot)$
is a functor from the category of B-branes to the category of left $\ZZ_2$-graded
modules over $Cl(n,\CC)$. Since spaces of open strings are expected
to be finite-dimensional, $M(Y)$ is expected to be a finite-dimensional
vector space. Thus $M(\cdot)$ is a graded functor from the graded category of B-branes
to $\ClMod(n)$. 

The results of Section~\ref{sec:D} (see also the Appendix) imply that the functor 
$M(\cdot)$ maps the maximal linear
B-brane to an irreducible Clifford module; for even $n$ maximal isotropic
subspaces which belong to different irreducible families are mapped to non-isomorphic
Clifford modules related by parity reversal. The D0-brane
is mapped to the free module of rank one. A linear B-brane of complex 
dimension $\ell< [n/2]$ is mapped to a module which is a direct sum of 
$2^{[n/2]-\ell}$ irreducible modules.

Next we would like to explain why $\ClMod(n)$ is equivalent to 
$\K(n)$.
The relation between these two rather different-looking categories is
a generalization of the so-called Koszul duality for quadratic 
algebras~\cite{Priddy,BGG,BGS}. 
Any serious attempt to discuss Koszul duality would take us out of our depth, 
so we will just make a few remarks which may help to orient the reader 
who would like to study these questions deeper. 

Classical Koszul duality applies to
quadratic algebras, i.e. $\ZZ$-graded algebras generated by 
degree-1 elements, such that all relations between generators are homogeneous
quadratic. The basic example of a dual pair is the pair $(S^*(V^*),\wedge^*V)$, where
$S^*(V^*)$ is the symmetric algebra of a finite-dimensional vector space $V^*$, and $\wedge^*V$
is the exterior algebra of the dual vector space. The statement of Koszul duality is that their derived categories
of finitely-generated $\ZZ$-graded modules are equivalent. 

There is a generalization of Koszul duality to the case 
where the relations are non-homogeneous quadratic~\cite{Priddy,P,PP}. But the
dual object in this case is not a graded algebra, but a quadratic CDG algebra.
A CDG algebra is a triple $(A,d,f)$, where $A$ is a graded algebra, $d$ is a degree-1 derivation,
and $f$ is a degree-2 element $f$ such that $d^2a=[f,a]$ for any $a\in A$.
CDG means ``curved differential graded''; another name for a CDG algebra
is a ``Q-algebra''~\cite{ASchwarz}. A module over a CDG algebra $(A,d,f)$ is a 
graded module $M$ over $A$ equipped with a degree-1 derivation $d_M$
such that $d_M^2 m=f\cdot m$ for any $m\in M$. 

What we need is a $\ZZ_2$-graded
version of non-homogeneous Koszul duality. Indeed, on one hand,
the Clifford algebra is a $\ZZ_2$-graded quadratic algebra, while on the other
hand, the category $C(X,W,W_0)$ can be regarded as a category
of modules over a certain $\ZZ_2$-graded CDG algebra. This CDG algebra
is purely even and isomorphic to $\O_X$ as an algebra.
The derivation $d$ is identically zero, but the even element $f$ is not:
it is given by $W$. The category $\K(n)$ can be regarded
as the derived category of the category of finitely generated CDG modules over the
CDG algebra $(\O_X,0,W)$. This CDG algebra is
Koszul-dual to the Clifford algebra in the sense of Refs.~\cite{P,PP}, and we expect that the corresponding derived
categories of modules are equivalent. 
More precisely, we expect that
the derived category of finite-dimensional $\ZZ_2$-graded Clifford modules 
is equivalent to the derived category of finitely-generated
modules over the CDG algebra $(\O_X,0,W)$.

Since the Clifford algebra can be regarded as a deformation
of the exterior algebra, and the CDG algebra $(\O_X,0,W)$ is a deformation of the polynomial algebra,
this claim looks like a generalization of the classic result of Ref.~\cite{BGG}. In fact, the deformed
duality is in some sense simpler than the classic one, since the category $\ClMod(n)$ is semi-simple and ``deriving''
it is a trivial operation (gives us back the same category). It is also more useful: while the classic duality of Ref.~\cite{BGG}
reduced the problem of classifying coherent sheaves on $\CC\PP^n$ to a {\it very difficult}
problem in linear algebra, the deformed duality reduces the problem of classifying B-branes
in the free massive LG model to a {\it very simple} problem in linear algebra (classification
of finite-dimensional graded modules over a Clifford algebra.)

Let us describe the functors which establish the equivalence of $\K(n)$ and $\ClMod(n)$.
The first one, from $\K(n)$ to $\ClMod(n)$, is obvious:
it takes an object $Y$ of $\K(n)$ to $Mor(Y_0,Y)$, where $Y_0$ is 
the object of $\K(n)$ described in the last paragraph of
subsection~\ref{sec:konts}. The mapping of morphisms is the obvious one.

To define the functor
acting in the opposite direction, let us consider for any object $M$
of $\ClMod(n)$ the vector space
$$
N=M\ot_\CC \O_X,
$$
where $\O_X$ is simply the algebra of polynomial functions on $\CC^n$. 
Since $M$ is $\ZZ_2$-graded, this vector space is also $\ZZ_2$-graded.
It is also an $\O_X$ module, for obvious reasons. It remains to define
the twisted differential $d_N$, i.e. an odd endomorphism of $N$ which squares to
$W$. Let $V$ be the vector space which appears in the definition of the
Clifford algebra; we will also identify the target space $X$ of the
LG model with $V$. The twisted differential will be
$$
d_N: m\ot f\mapsto \sum_i (e_i\cdot m)\ot z_i f,\quad \forall m\in M, \quad\forall f\in \O_X,
$$
where $e_i,i=1,\ldots,n,$ is a basis in $V$, $z_i$ are the corresponding
linear coordinates, and the dot denotes the Clifford algebra action.
It is easy to check that $d_N$ is odd, and that $d_N^2=W$. Thus we defined a map
which sends an object of $\ClMod(n)$ to an object of $\K(n)$. The mapping of morphisms
is the obvious one: if $\alpha$ is a morphism of Clifford modules $M$ and $M'$, then
the corresponding element of ${\rm Hom}_{\O_X}(N,N')$ is $\alpha\ot 1$.
It is easy to check that $\alpha\ot 1$ is closed, and thus is a well-defined morphism in the
category $\K(n)$.

The claim is that compositions of these two functors in any order
are isomorphic to identity functors. We will not try to prove
this claim here, but to make it more plausible note that the mapping of objects is given
by essentially the same formulas as in the classic case~\cite{BGG}.

\section{Application: the category of A-branes for some Fano varieties}
\label{sec:F}

\subsection{A-branes on $\CC\PP^2$}

The mirror of the the nonlinear sigma model with target $\CC\PP^2$ is the affine $A_2$ Toda model~\cite{HV}.
The affine $A_2$ Toda model is an $\N=2$ \LG theory of two chiral
superfields $x$ and $y$ taking values in $\CC^*$ and a rational superpotential $$W(x,y) =
x+y+\frac1{xy}.$$ We can test the Homological Mirror Symmetry conjecture by comparing the Fukaya category
of $\CC\PP^2$ with the category of B-branes in the Toda model.

\begin{figure}[b]
  \centerline{\epsfxsize=10cm\epsfbox{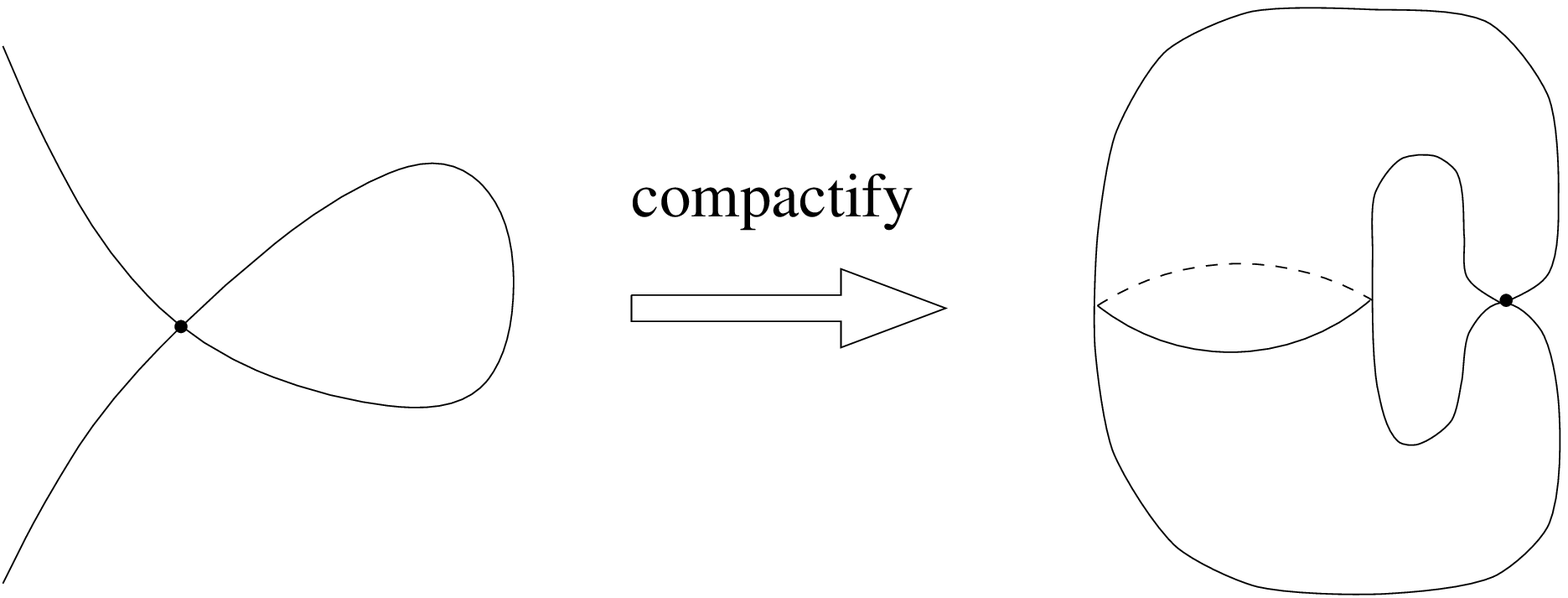}}
  \caption{}
  \label{fig:cubic}
\end{figure}

As discussed above, every B-brane in the Toda model lies on some holomorphic curve
$W=W_0$.  In addition, in order for open strings to have a supersymmetric ground state, we require
this curve to pass through a critical point of $W$. In the $A_2$ theory there are three distinct
critical points:  $$x=y=a_k=e^{2k\pi i/3},\qquad k=0,1,2.$$ The values of $W$ corresponding to
these critical points are pairwise distinct: $W_k=3a_k$. There is an obvious $\ZZ_3$ symmetry which permutes
the critical points. This implies that the categories $H^*(C(X,W,W_k))$ are all equivalent.
{}From now on we will focus on one of them, say, the one corresponding
to $k=0$. All B-branes associated to this critical point must be complex submanifolds of the
holomorphic curve in $\CC^*\times \CC^*$ given by
\begin{equation}\label{curve}
x+y+\frac{1}{xy}-3=0.
\end{equation}
This curve is a singular cubic with a single node (see Fig.~\ref{fig:cubic}). Thus the category of B-branes is a full sub-category
of the category of B-branes in the LG model $W=xy$. We have seen that the latter is equivalent to the
category $\ClMod(2)$. 

It remains to understand which objects in the latter category correspond to B-branes. Clearly, the D0-brane
sitting at the critical point $(1,1)$ is a valid B-brane.
As for D2-branes, they
must be (desingularizations of the) irreducible components of the curve Eq.~(\ref{curve}). But it is easy to see that the
singular cubic is irreducible. Thus there is only one D2-brane of type B:
the one which corresponds to the structure sheaf of the desingularized cubic. The corresponding
object in the ``local'' category associated to the critical point is the direct sum of the D2-brane
$x=0$ and the D2-brane $y=0$. This direct sum is isomorphic to the D0-brane (see Section~(\ref{sec:Cp})).
We conclude that the basic B-brane in the Toda model is the D0-brane, all other branes being direct
sums of several copies of the D0-brane. The endomorphism algebra of the D0-brane is isomorphic to
$Cl(2,\CC)$. We see that the category of B-branes in this case is strictly smaller than the
``local'' category $\C_{tot},$ which is equivalent to $\ClMod(2)$.

As discussed in Section~\ref{sec:B}, since the D0-brane looks like a composite of two D2-branes,
one can formally add these missing D2-branes to the category of B-branes for the Toda
model. The enlarged category is equivalent to the category $\ClMod(2)$.

Now let us interpret these results from the point of view of Homological Mirror Symmetry. 
The mirror of the D0-brane has been identified in Ref.~\cite{H} using the
dualization argument of Ref.~\cite{HV}. The mirror is a certain Lagrangian 2-torus in $\CC\PP^2$ 
equipped with a rank one trivial vector bundle and a certain flat connection. 
Let us be more specific.
Consider the unit 5-sphere in $\CC^3$, i.e. a hypersurface defined by the equation
$$
|z_1|^2+|z_2|^2+|z_3|^2=1.
$$ 
The quotient of this 5-sphere by a free $\SS^1$ action
$$
z_i\ra e^{2\pi i\alpha} z_i,\quad  i=1,2,3,\quad \alpha\in \RR/\ZZ,
$$
is diffeomorphic to $\CC\PP^2$. In fact, the standard symplectic form on $\CC\PP^2$ is obtained by restricting to $\SS^5$ the
standard K\"ahler form on $\CC^3$ and then pushing it down to the quotient. Now consider a 3-torus in $\CC^3$ defined
by the equations
\begin{equation}\label{3torus}
|z_1|^2=|z_2|^2=|z_3|^2=\frac{1}{3}.
\end{equation}
It is contained in the 5-sphere and invariant with respect to the $\SS^1$ action. Hence by passing to the quotient,
we obtain a 2-torus embedded in $\CC\PP^2$. It is trivial to check that this 2-torus is Lagrangian with respect
to the standard symplectic form on $\CC\PP^2$. The flat connection can be specified by its monodromy representation.
Let $\gamma_1$ and $\gamma_2$ be the loops on the 3-torus~(\ref{3torus}) defined by
$$
\gamma_1: t\mapsto \left\{\frac{1}{\sqrt 3}e^{2\pi i t},\frac{1}{\sqrt 3},\frac{1}{\sqrt 3}\right\},\quad
\gamma_2: t \mapsto \left\{\frac{1}{\sqrt 3},\frac{1}{\sqrt 3}e^{2\pi i t},\frac{1}{\sqrt 3}\right\}.
$$
Their images under the quotient map generate the fundamental group of our Lagrangian 2-torus. According to Ref.~\cite{H},
the mirror of the D0-brane sitting at the point $(a_k,a_k),$ $k=0,1,2,$ corresponds to the monodromy representation
which maps both generators to $e^{2\pi ik/3}$. In particular, the D0-brane which sits at the point $(1,1)$ is mirror
to the trivial flat connection on the Lagrangian 2-torus. 

As a simple check of this claim, 
note that the algebra of endomorphisms of a D0-brane in the model $W=xy$ has Euler characteristic zero. 
In the mirror picture, the corresponding object is the Euler characteristic of the Floer complex, which coincides
with the classical Euler characteristic of the 2-torus. Thus the Euler characteristics match. It would be nice to 
compute the Floer homology groups as well and to check that they agree with the predictions of mirror symmetry.
Namely, we expect that
\renewcommand{\theenumi}{\roman{enumi}}
\begin{enumerate}
\item the Floer homology of the Lagrangian 2-torus equipped with a rank-one flat connection is non-vanishing only for 
the three special flat connections defined above;
\item for these choices of the flat connection, the Floer homology is isomorphic to the classical cohomology of the torus as a 
$\ZZ_2$-graded vector space;
\item as a $\ZZ_2$-graded algebra, the Floer homology is isomorphic to the Clifford algebra with two generators, 
i.e. it is a quantum deformation of
the classical cohomology ring;
\item Floer homology groups which compute morphisms between different flat connections of rank one vanish.
\end{enumerate}

It was argued above that we can formally add D2-branes to the category of B-branes. It is reasonable to ask if this procedure
is consistent with or perhaps even forced on us by Homological Mirror Symmetry. To answer this question we need to identify the mirrors of
the added D2-branes. There are two such D2-branes for each critical level set.  For each of them the Euler characteristic of 
the endomorphism algebra is $1$. If we assume that the mirror of a D2-brane is a Lagrangian submanifold, then it must be 
homeomorphic to a real projective
plane $\RR\PP^2$. But since $\RR\PP^2$ is not orientable, it is not an admissible object of the Fukaya category
(one needs orientability in order to define $\ZZ_2$-graded Maslov index and $\ZZ_2$-grading on the Floer complex).
We conclude that the mirrors of the added D2-brane cannot be Lagrangian submanifolds, and therefore Homological
Mirror Symmetry does not force us to include them on the B-side.

On the other other hand, if we added D2-branes on the B-side, we can maintain Homological Mirror Symmetry by adding certain objects
on the A-side. In other words, we would like to regard the Lagrangian 2-torus with a trivial flat connection, which is mirror to
the D0-brane, as a direct sum of two irreducible objects, which are mirror to the D2-branes. But since there are no such objects in 
the Fukaya category, we simply add these direct summands ``by hand.''

Let us clarify what we mean by adding direct summands ``by hand.'' Let $E$ be an object of an additive 
category ${\mathsf C}$. A projector is an element of $\End(E)$ which
satisfies $e\circ e=e$. Given any projector, we would like to have the corresponding direct summand, i.e. an object $R$ and
a pair of morphisms $i:R\ra E$ and $r:E\ra R$ such that $r\circ i=id_R$ and $i\circ r=e$. If $R$ does not exist for all projectors
and for all $E$, then we look for the smallest additive category which contains ${\mathsf C}$ as a full subcategory and in which every 
projector has a direct summand. 

To summarize, to maintain Homological Mirror Symmetry, we must either add formal direct summands on both A and B sides, or on neither side.

\subsection{A-branes on $\CC\PP^1\times\CC\PP^1$}

The mirror in this case is the LG model with target $\CC^*\times\CC^*$ and the superpotential 
$$
W(x,y)=x+\frac{\mu}{x}+y+\frac{\nu}{y}.
$$ 
Here $\mu$ and $\nu$
are nonzero complex numbers whose logarithms are mirror to the periods of the complexified K\"ahler form on the
two $\CC\PP^1$'s. This superpotential has four non-degenerate critical points. For generic $\mu,\nu$ there are four critical level
sets all of which look like a cubic with a node. Thus we are in exactly the same situation as in the previous subsection,
and the only B-branes are D0-branes sitting at the critical points. Another way to see these D0-branes is to note
that the LG model is a product of two LG models with the superpotential $W=x+\mu/x$. This model is mirror to
$\CC\PP^1$ and has been studied in Ref.~\cite{H}. Its only B-branes are D0-branes sitting at the two critical points
of the superpotential. Taking tensor products of pairs of such B-branes gives us four D0-branes discussed above.

The mirror of each D0-brane is a Lagrangian 2-torus with some flat connection. Indeed, the mirror of a D0-brane
in the model $W=x+\mu/x$ is the equatorial circle on $\CC\PP^1$~\cite{H}, therefore the mirror of a D0-brane in the
product model is the product of two equatorial circles. The monodromy around the two generators of the fundamental
group is $(\pm 1,\pm 1)$.

For $\mu=\nu$ something special happens both on the A and B sides. On the A side, we get a new Lagrangian submanifold
which is homeomorphic to a 2-sphere. To see this, let $z$ and $w$ be coordinates on the standard affine patches
on the two $\CC\PP^1$'s. Consider the ``anti-diagonal'' 2-sphere given by $z=\bar{w}$. Let $\omega$ be the
Fubini-Study form on $\CC\PP^1$, $\pi_i, i=1,2,$ be the projection maps from $\CC\PP^1\times\CC\PP^1$ to the two factors,
and $a_1,a_2$ be complex numbers.
It is trivial to check the restriction of $a_1\pi_1^*\omega+a_2\pi_2^*\omega$ to the ``anti-diagonal'' 2-sphere
vanishes if and only if $a_1=a_2$. Thus the 2-sphere is Lagrangian if and only if $\mu=\nu$.

On the B side, the critical level set $W=0$ now contains two critical points. The equation of this critical level set
$$
(x+y)(xy+\mu)=0
$$
shows that it is reducible. The irreducible components are a line and a non-singular quadric which
intersect transversally at two points; these are the two critical points mentioned above. We have two irreducible
D2-branes of type B corresponding to the two irreducible components of the critical level set. It is easy to see that one 
is isomorphic to the shift of the other, while their sum is isomorphic to the sum of two D0-branes sitting at
the two critical points.

Note that this is another example where the category of B-branes is strictly smaller than the sum of ``local'' categories
associated to critical points. This happens because all D2-branes pass through both critical points in the set $W=0$. 
Thus a single D0-brane sitting at a critical point is irreducible. Of course, if we only look at
the infinitesimal neighborhood of one of the critical points, then we are in the same situation as in the model
$W=xy$, and the D0-brane appears to be composite. If desired, we can enlarge the category of B-branes by adding
all formal direct summands. Then it will become equivalent to the sum of categories attached to the two critical points (each
of which is equivalent to $\ClMod(2)$), and each D0-brane will be the sum of two irreducible objects.

Now let us match the objects on A and B sides. D0-branes correspond to ``equatorial'' Lagrangian tori, as before.
The mirror of a D2-brane must be a Lagrangian 2-sphere. Indeed, each D2-brane passes through two critical points,
each of which contributes $1$ to the Euler characteristic of the endomorphism algebra. An obvious conjecture is that
the two D2-branes are mirror to the Lagrangian 2-sphere discussed above and its shift (i.e. orientation-reversal).
If this is true, then the sum of the Lagrangian 2-sphere and its shift must be isomorphic 
(in the Fukaya category) to the sum of two
``equatorial'' Lagrangian tori with monodromies $(1,-1)$ and $(-1,1)$. It would be interesting to check this by computing
the Floer homology between all the objects involved.

$$$$

\vspace{-5mm}

\section{Comments and Outlook}\label{sec:G}

In this paper we have described the category of B-branes for the free
massive LG model with $n$ chiral fields. We also argued
that this allows one to determine the category of B-branes for an 
arbitrary massive LG model. The most striking feature of our results
is their simplicity. For example, if we consider the free massive LG model, 
there is a multi-parameter family of maximal isotropic subspaces of the
quadric $W=0$, but they are all isomorphic as objects
of the category of B-branes (up to a shift). Moreover, B-branes of lower
dimension, including the D0-brane, are isomorphic to direct sums of
B-branes of maximal dimension. These rather counter-intuitive observations solve
the problem of computing tree-level topological open string correlators in these models.

It is interesting to compare our results
with those of Refs.~\cite{MS,Laz}, where a general framework for classifying
D-branes in 2d Topological Field Theories has been proposed. In our
case, the 2d TFT in the bulk is rather trivial: it is isomorphic, as a 
Frobenius algebra, to $\CC$ with its unique Frobenius structure. The theory
of Ref.~\cite{MS} (generalized to the $\ZZ_2$-graded case) tells us
that the algebra of open strings connecting a brane with itself must
be simple. We saw that in our case endomorphism algebras
of B-branes are all isomorphic to Clifford algebras, and these are indeed
simple (as $\ZZ_2$-graded algebras). However, unlike in the purely
bosonic case, in the $\ZZ_2$-graded case not every two simple finite-dimensional
algebras are Morita equivalent. In fact, there are two Morita-equivalence
classes of such algebras, represented by $\CC$ and $Cl(1,\CC)$. A
Clifford algebra with $k$ generators is Morita-equivalent to $\CC$
or $Cl(1,\CC)$ depending on whether $k$ is even or odd. We have seen that
when the number of fields $n$ is even (resp. odd) only even (resp. odd)
values of $k$ occur. This suggests that it is impossible to have
a topological open string theory which includes D-branes of both kinds.
Indeed, we have seen that all pairings between spaces of morphisms induced
by 2-point correlators
are either even or odd, depending on whether $n$ is even or odd. On the other hand,
one of the basic axioms of topological open string theory is that all
pairings must have the same parity~\cite{Laz}.

This observation provides a simple counter-example to the belief that
a 2d TFT determines uniquely the associated category of topological boundary
conditions. In fact, we can make a stronger statement. Given any 2d SCFT
representing a superstring background,
we can tensor it with the topological LG model $W=z^2$. Since the latter
theory is trivial, this does not change the closed string sector.
But the open string sector does change: one has to tensor every ``physical''
D-brane with the D0-brane of the LG model, and this results in tensoring
the open string spectrum of each D-brane with $Cl(1,\CC)$. This is equivalent
to the introduction of an odd Chan-Paton label. Thus we have two inequivalent
open string theories for a given closed string theory. 

This particular ambiguity is rather mild and can be easily eliminated.
The difference between odd and even $n$ comes from the number of fermionic
zero modes on a disk, or equivalently from the parity of the bilinear forms
computed by the 2-point disk correlators. Thus to specify completely the open string theory we are
dealing with, it is sufficient to fix the parity of all bilinear forms.

It would be interesting to extend the considerations of this
paper to LG models which flow to non-trivial SCFTs in the infrared limit.
For example, one could study Landau-Ginzburg realizations of $N=2$
minimal models. For these theories much information about B-branes
is available from the boundary state formalism, and it would interesting
to see if it is consistent with Kontsevich's proposal.
By analogy with the massive case, one expects that the category
of B-branes will be describable in terms of modules over the algebra
of endomorphisms of a D0-brane. From the mathematical viewpoint,
this algebra must be related by a Koszul-like duality to the CDG algebra
which appears in Kontsevich's proposal. It appears that for $W$ of degree
higher than two Koszul duality relates Kontsevich's CDG algebra to
a finite-dimensional $A_\infty$-algebra~\cite{Kpriv}. B-branes should correspond
to finite-dimensional $A_\infty$-modules over this $A_\infty$-algebra. 
In this way solving topological open string theory is
reduced to a problem in linear algebra. Hopefully, the latter problem is
manageable. 

In the axiomatic approach of Refs.~\cite{Laz,MS}, topologically twisted
$N=2$ minimal models correspond to non-semi-simple Frobenius algebras.
It would be interesting to explore the uniqueness of the open string sector
in such models.  

\section*{Appendix: Clifford algebras and modules}

In this appendix we collect some well-known facts about complex Clifford
algebras and their modules. Let $V$ be a complex vector space of dimension
$n$, and $Q$ be a non-degenerate symmetric bilinear form on $V$.
Clifford algebra $Cl(V,Q)$ has $V$ and the identity as its set of generators, and the
following relations:
$$
v\cdot v'+ v'\cdot v=Q(v,v').
$$
As a vector space, $Cl(V,Q)$ is isomorphic to $\wedge^*V$ and therefore
has dimension $2^n$. We can regard $Cl(V,Q)$ either as an ordinary 
associative algebra, or as a $\ZZ_2$-graded algebra, such that all
the generators are odd. In the latter case, the grading corresponds
to the decomposition of $\wedge^*V$ into polyvectors of even and odd
degree. Since the isomorphism class of $Cl(V,Q)$ depends only on the
dimension $n$ of $V$, we will also use the notation $Cl(n,\CC)$ to denote
this isomorphism class.

If $V_1$ and $V_2$ are complex vector spaces with non-degenerate bilinear
forms $Q_1$ and $Q_2$, then 
\begin{equation}\label{tensorCl}
Cl(V_1\op V_2,Q_1\op Q_2)=Cl(V_1,Q_1)\ot Cl(V_2,Q_2).
\end{equation}
Here all Clifford algebras are regarded as $\ZZ_2$-graded algebras,
and $\ot$ denotes their $\ZZ_2$-graded tensor product.

If $n$ is even, then $Cl(V,Q)$ regarded as an ungraded algebra is
isomorphic to the algebra of complex $2^{n/2}\times 2^{n/2}$ matrices,
which we will denote $Mat(2^{n/2},\CC)$. If $n$ is odd, then 
$Cl(V,Q)$ regarded as an ungraded algebra is isomorphic to
$Mat(2^{[n/2]},\CC)\op Mat(2^{[n/2]},\CC)$. In particular, $Cl(1,\CC)$
is isomorphic to $\CC\op\CC$. We see that $Cl(V,Q)$
is a simple algebra only for even $n$. However, if we regard it as
a $\ZZ_2$-graded algebra, then it is simple for all $n$. 

Now let us discuss finite-dimensional modules over $Cl(V,Q)$. The
category of Clifford modules is semi-simple, i.e. every exact sequence
splits. Thus every Clifford module is a direct sum of irreducible modules.
The number and properties of irreducible modules depend on the parity
of $n$, as well as whether we regard $Cl(V,Q)$ as a $\ZZ_2$-graded algebra.
If we neglect the grading, then for even $n$ we have a unique irreducible
module $S$ of dimension $2^{n/2}$. It is called the spinor module
and can be constructed as follows. Pick a pair of subspaces $U,W$ of $V$ 
such that both $U$ and $W$ are isotropic with respect to $Q$,
$U\bigcap W=0$, and $V=U\op W$. 
One can easily see that $Q$ gives a non-degenerate pairing between $U$ and
$W$ and thus we may identify $W$ with $U^*$.
Set $S=\wedge^* U$, and define the action of Clifford algebra on $S$
as follows: if $v=u\op w,$ where $u\in U$ and $w\in W$, 
then for any $\lambda\in S$ we let
$$
(u\op w)\cdot \lambda=u\wedge \lambda + i_w\lambda.
$$
Here we used the identification of $W$ with $U^*$ mentioned above.

For odd $n$ Clifford algebra is a sum of two matrix algebras, and 
therefore there are two non-isomorphic irreducible modules of dimension
$2^{[n/2]}$ (two spinor modules). For example, for $n=1$ the algebra is 
generated by the identity and an odd element $\xi$ with a single relation $\xi^2=1$; 
the two irreducible modules are one-dimensional, with the action of $\xi$ 
given by $\pm 1$. For general $n$ one can use the property Eq.~(\ref{tensorCl})
to reduce the problem to the cases already considered.

If we regard $Cl(V,Q)$ as a $\ZZ_2$-graded algebra, then we should
look for $\ZZ_2$-graded irreducible modules. For even $n$ there are
two inequivalent choices of grading on the spinor module related by parity
reversal. Therefore we have two non-isomorphic irreducible spinor modules
$S$ and $\bar{S}$. For odd $n$ the ``minimal'' $\ZZ_2$-graded module
has dimension $2^{(n+1)/2}$; as an ungraded module, it is isomorphic
to the direct sum of two inequivalent irreducible ungraded modules. Furthermore, the
choice of grading is unique up to isomorphism. We will denote this
unique spinor module by $S$. For example, for $n=1$ $S\simeq \CC^2$,
and $\xi$ acts as any of the three Pauli matrices, say $\sigma_1$.
Then the parity operator can be chosen to be $\sigma_3$. 

To summarize, for even $n$ any $\ZZ_2$-graded Clifford module is a direct
sum of several copies of two inequivalent spinor modules $S$ and $\bar{S}$.
For odd $n$ the situation is the same, except that $S$ is isomorphic to
$\bar{S}$. The dimension of the spinor module is given by $2^{[(n+1)/2]}$
for any $n>0$.

In particular, $Cl(V,Q)$ regarded as a left module over itself is a direct
sum of $2^{[n/2]}$ copies of spinor modules. For $n$ even half of them
are $S$'s, and the other half are $\bar{S}$'s.

\section*{Acknowledgments}

We are deeply grateful to Maxim Kontsevich for sharing with us his ideas about
B-branes in Landau-Ginzburg models, and to Alexander Polishchuk for pointing out the
relevance of non-homogeneous Koszul duality.
We also thank Kentaro Hori and Dmitri Orlov for reading a preliminary draft of the paper and making a 
number of valuable comments. The first author would like to thank Institut des Hautes \'{E}tudes Scientifiques for
hospitality during the writing of this paper. This work was supported in part by the DOE grant DE-FG03-92-ER40701.

\end{document}